# Asteroseismically Inferred Ages of 132,000 Red Giants with TESS

Artemis Theano Theodoridis, Leslie Morales, and Jamie Tayar
Department of Astronomy, University of Florida, FL, USA

## Abstract

NASA's Transiting Exoplanet Survey Satellite mission has identified at least 158,000 oscillating red giants, increasing the known sample by roughly an order of magnitude. After validating that these measurements are reliable to 5% for up to 90% of red giants, we make custom stellar evolution models using Modules for Experiments in Stellar Astrophysics in order to estimate ages for ∼132,794 of these stars to an average uncertainty of 23%. We show that these ages follow similar distributions to those observed in other samples such as Kepler, with small differences likely resulting in the Galactic volume probed. We provide these ages to the community to enable future Galactic archeology analyses.

## 1. Introduction

Accurate calculations of stellar ages are integral to understanding the formation and evolution of our Galaxy. Ages have been used to trace the evolution of the Galactic bulge (M. Joyce et al. 2023), the thin and thick disks (A. Helmi et al. 2018; D. K. Feuillet et al. 2022), the halo (G. Iorio et al. 2018; J. T. Mackereth et al. 2018), accreted dwarf galaxies (D. K. Feuillet et al. 2022; S. K. Grunblatt et al. 2022), as well as interactions between these components (W. J. Chaplin et al. 2020). These Galactic archeology studies have determined that the Galaxy has different components and that they have evolved and interacted with each other over time. Having ages for these stars allows us to understand the chronological order of their evolution.

However, there are still two major challenges in this field: (1) In past decades, precise and accurate age interpolation and determination have been limited to small portions of the sky, which may not be generalizable to the full Galaxy. (2) There are also disagreements on the age of each star depending on the model and methods chosen to derive said ages, which can impact the inferred Galactic history (e.g., J. Tayar & M. Joyce 2025). These limitations lead to the debate over the accuracy of individual studies and significant differences in derived ages for larger samples (e.g., comparing V. Silva Aguirre et al. 2018; M. Xiang & H.-W. Rix 2022; A. Stokholm et al. 2023; D. M. Nataf et al. 2024).

Since ages are not a directly measurable property in most cases, almost all ages are derived in reference to a stellar model. However, numerous physical assumptions must be made to construct a stellar model (M. Joyce et al. 2024) and many can impact the inferred age (J. M. Ying et al. 2023). In most commonly used models (G. Takeda et al. 2007), physics such as convective overshoot, helium abundances, atmospheric boundary conditions, and mixing length are heavily simplified. This can lead to different ages across studies for the same star depending on the stellar models used. Various studies explore the complexities of stellar modeling physics alongside their uncertainties (J. L. van Saders & M. H. Pinsonneault 2013; J. Choi et al. 2016; M. Joyce & B. Chaboyer 2018), and in general they find that much of the physics of stellar evolution is still truly uncertain at some significant level. So even though individual authors are all making reasonable assumptions, they may predict slightly different evolution for the same star. While addressing these complex physical processes is a long-term ongoing effort, there are a significant number of cases where models agree quite well and can be used to draw interesting conclusions about the history of the Galaxy.

In particular, work has been done to map the differences in inferred ages from different model grids (J. Tayar et al. 2022; S. Byrom & J. Tayar 2024; L. Morales et al. 2025). In general, ages can be uncertain to 20% for main-sequence and subgiant stars and up to 80% for red giants when using isochrones or stellar models to fit ages with well-constrained photometric or spectroscopic values (J. Tayar et al. 2022; L. Morales et al. 2025). However, the age of a star is strongly related to its mass, and so when the mass of a star is directly measured or calculated, the discrepancy between models drops significantly. This is particularly relevant for red giant branch (RGB) stars (L. Morales et al. 2025; M. H. Pinsonneault et al. 2025), where small changes in mass can lead to significant differences in their inferred age. The large discrepancies between models when mass is unknown arise from the fact that models of different masses have very similar temperatures on the RGB, and so a slight shift in the model temperature can change the inferred mass significantly. Therefore, when estimating ages for samples of red giant stars, it is strongly preferable to choose stars where the stellar mass can be inferred directly. One of the best ways to get masses for large samples of single stars is asteroseismology.

Asteroseismology, the study of stellar oscillations, provides further knowledge of a star's internal structure and evolution (T. M. Brown et al. 1991; H. Kjeldsen & T. R. Bedding 1995; R. L. Gilliland et al. 2010; W. J. Chaplin & A. Miglio 2013). Oscillations within stars excite certain frequencies that are related to the bulk properties of the star. For stars with a solar-like structure, that is, a convective surface, turbulent convective motions will excite waves for which pressure is the restoring force. In solar-like oscillators, these modes are excited across a range of frequencies near $\nu_{\max}$, the frequency of maximum power, which scales with the star's surface gravity (T. M. Brown et al. 1991). The power envelope is approximately Gaussian, but the underlying oscillations can include pressure ($p$ modes) and mixed modes, the latter of which can become important in evolved stars such as red giants (T. R. Bedding et al. 2011). Mode excitation is primarily determined by convective driving near $\nu_{\max}$ (e.g., W. J. Chaplin & A. Miglio 2013), rather than purely by the restoring forces of individual modes. There are regular spacings between modes of consecutive radial order $n$, for the same spherical degree $\ell$, which are called the large frequency spacing, or $\Delta\nu$, and this will scale with the mean density. Combining various stellar components such as $\nu_{\max}$, $\Delta\nu$, and effective temperature allows for a calculation of the star's mass, which, in combination with a metallicity and a stellar model, can provide a more precise estimate of a star's age (R. K. Ulrich 1986).

It was only recently that the Kepler mission made it possible to use this technique for moderate samples (i.e., tens of thousands of stars). Galactic archeology goals need truly large datasets, more like hundreds of thousands or millions of stars. Many researchers have used machine learning to bridge the gap (M. Ness et al. 2016; H. W. Leung & J. Bovy 2019; J. T. Mackereth et al. 2019; F. Anders et al. 2023; A. B. A. Queiroz et al. 2023; A. Stone-Martinez et al. 2024) but this approach has some challenges. In particular, there are often difficulties with interpretability, identifying adequate and unbiased training sets, and mitigating edge-case inaccuracies (J. Tayar et al. 2023; Y.-S. Ting 2025). For both direct asteroseismic analysis and identifying training samples for machine learning approaches, NASA's Kepler (W. J. Borucki et al. 2010) is almost unanimously considered the gold standard for asteroseismic samples because of its 4 yr long continuous observation of the same part of the sky, allowing detailed measurements of frequencies that are well resolved in the Fourier domain. That said, this sample is somewhat limited by its small field of view and the specific population of stars available in that portion of the sky. Despite the relatively small sample size, incredible advancements have been made in the study of Galactic archeology using machine learning trained on this small sample of a few thousand stars. For example, nearly 100,000 stellar ages have been inferred using machine learning methods trained on asteroseismic results from large spectroscopic surveys, and detailed maps have been made of the evolutionary history of the Milky Way (M. Ness et al. 2016; H. W. Leung & J. Bovy 2019; F. Anders et al. 2023; A. B. A. Queiroz et al. 2023; A. Stone-Martinez et al. 2024).

Although these measurements have significantly advanced the field of Galactic archeology, NASA's Transiting Exoplanet Survey Satellite (TESS; G. R. Ricker et al. 2010) has the potential to drastically increase the number of asteroseismic ages available. TESS is an all-sky survey that includes millions of observed red giants (J. T. Mackereth et al. 2021; E. J. Hatt et al. 2024; J. Zhou et al. 2024), 158,000 of which are confirmed identified oscillators (M. Hon et al. 2021). In the M. Hon et al. (2021) article specifically, machine learning was used to identify potential red giant oscillators and estimate a frequency of maximum power. Combining that $\nu_{\max}$ with a stellar radius estimate from Gaia and either a photometric or spectroscopic temperature estimate is sufficient to compute a seismically informed mass. Previous work (A. T. Theodoridis & J. Tayar 2023) has shown that TESS seismology is calibrated to the Kepler scale to better than 5% for about 90% of red giants, indicating an excellent agreement to data that are considered a gold standard and capability for future studies. This allows us to address the two problems mentioned earlier: (1) We are no longer limited to a small field of view in our data since TESS has looked at almost the whole sky over the course of its mission. (2) These ages come from stars of constrained mass, which reduces the model-dependent systematics in red giant ages from ~80% to ~10% (L. Morales et al. 2025).

These advancements allow us to provide a more direct age estimate for a very large sample of stars, which can then be used for Galactic archeology studies, contributing to our understanding of the formation and evolution of the Milky Way and its red giant stellar populations. We also hope that this sample can be used to identify interesting stars or subsample for more detailed individual studies of interesting objects in the future, although we do not attempt such studies here.

## 2. Data

While it is in some cases possible to estimate the ages of stars from their temperatures, luminosities, and metallicities (D. Godoy-Rivera et al. 2021; M. Xiang & H.-W. Rix 2022; D. M. Nataf et al. 2024, etc.) work done by L. Morales et al. (2025) has shown that such ages for giant stars depend significantly on the stellar models used. However, ages estimated from the stellar mass and metallicity are much less sensitive to the chosen model. Therefore, we define a set of stars where we can estimate mass, metallicity, and surface gravities on the RGB from a combination of TESS and Gaia data.

### *2.1. TESS Seismology*

Work done by M. Hon et al. (2021) identified 158,000 oscillating red giants from TESS data, which we focus on in this study. The construction of the TESS catalog utilized deep learning, a form of supervised machine learning, which was initially applied to Kepler and K2 data to detect red giant oscillators and estimate $\nu_{\max}$ values with performance comparable to human experts (M. Hon et al. 2018). This technique involved two independently trained convolutional neural networks that analyzed 2D images of power spectra, effectively "visually" identifying red giant oscillation signatures.

This method was applied to TESS data, analyzing long-cadence photometry taken at 30 minute intervals for 1 month (27 days). Using data from the MIT Quick-Look Pipeline (C. X. Huang et al. 2020a, 2020b),[1] which extracts light curves from full-frame images, the analysis led to the identification of oscillations in approximately 158,000 red giants over the TESS 2 yr primary mission (Years 1 and 2, Cycles 1–26).

The best seismology available to date (e.g., M. N. Lund et al. 2017; J. C. Zinn et al. 2019; J. Schonhut-Stasik et al. 2023; M. H. Pinsonneault et al. 2025) comes from extensive datasets with long observation times and careful analysis by many researchers. For this reason, the use of machine learning on short datasets has faced skepticism. However, the TESS data have been validated through comparison with independent asteroseismic pipelines in well-studied regions, including stars that were independently observed by both the TESS and Kepler missions (D. Stello et al. 2022). TESS studies in the Southern Continuous Viewing Zone using multiple techniques were also found to be in good agreement with each other (J. T. Mackereth et al. 2021), suggesting that the TESS data in themselves are of high quality for asteroseismic analyses.

[1] https://archive.stsci.edu/hlsp/qlp

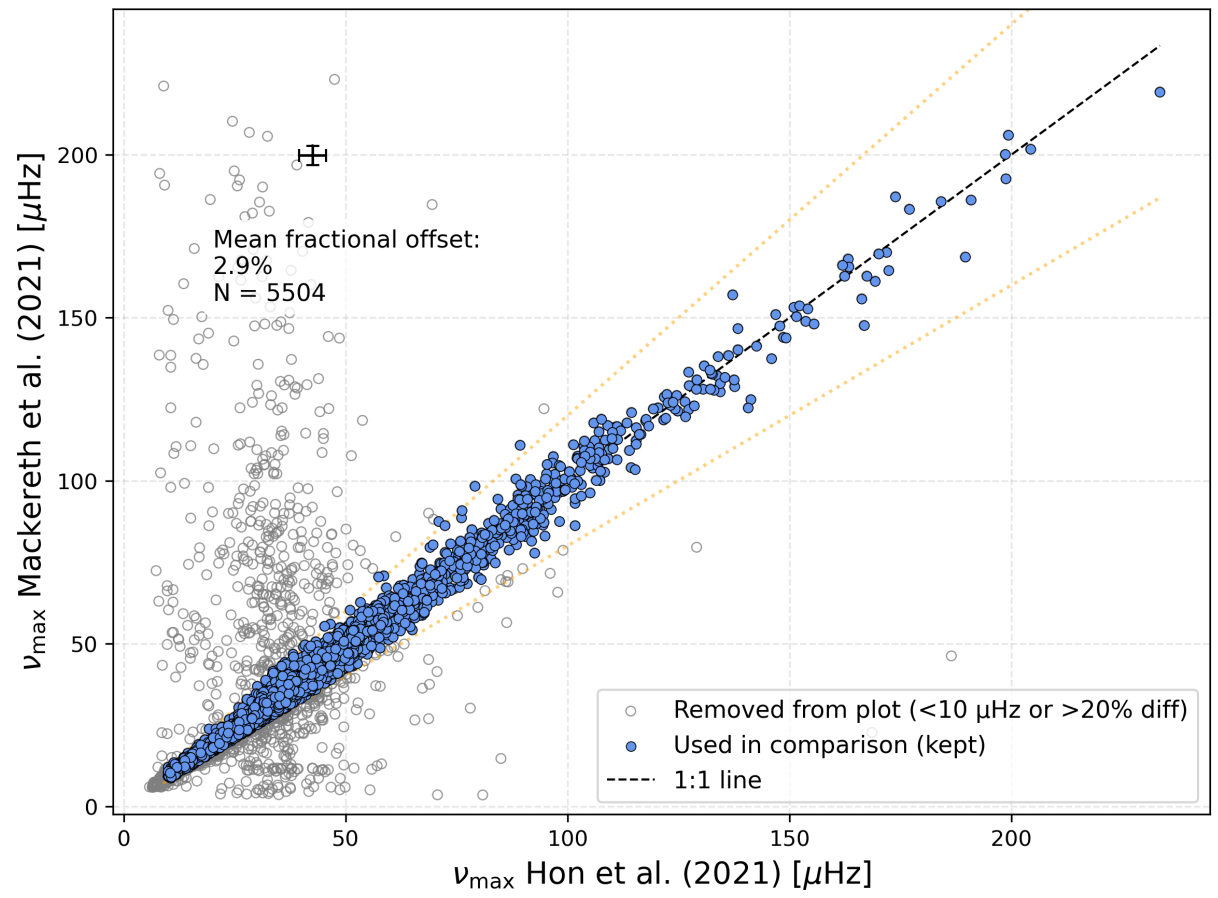

**Figure 1.** Comparison of $\nu_{\rm max}$ values from M. Hon et al. (2021) and J. T. Mackereth et al. (2021). Gray open circles indicate stars excluded from the comparison (either with $\nu_{\rm max} < 10\ \mu$Hz or differing by more than 20%), while blue points show those retained for analysis. These stars are only excluded from this plot, not from the overall sample. After removing outliers, we find a mean fractional offset of 2.9% with a standard deviation of 6%, demonstrating good agreement between the two catalogs. For subsequent analyses, we adopt a conservative uncertainty of ∼6% on the $\nu_{\rm max}$ values from M. Hon et al. (2021).

Additionally, machine learning seismic $\nu_{\rm max}$ values were found to be consistent with spectroscopically calibrated $\nu_{\rm max}$ values on the Kepler scale (A. T. Theodoridis & J. Tayar 2023). Across various studies, our $\nu_{\rm max}$ values show consistent agreement with independent analyses: (1) Kepler-to-TESS comparisons (D. Stello et al. 2022) validate the general reliability of TESS data, (2) multiple TESS pipeline reductions (J. T. Mackereth et al. 2021; E. Hatt et al. 2022) yield consistent results (∼1%–2% scatter), and (3) machine learning estimates agree with spectroscopy (A. T. Theodoridis & J. Tayar 2023) that has been calibrated to the Kepler scale. For stars with overlapping analyses (M. Hon et al. 2021), sector-to-sector variations are minimal (<3% vary by more than 2$\sigma$), and interstudy comparisons show comparable precision. While most verifications use TESS data, the consistency across different methodologies and the Kepler–TESS cross validation suggest our $\nu_{\rm max}$ scale is robust, with uncertainties typically within the quoted errors (J. Zhou et al. 2024). We also compared $\nu_{\rm max}$ values from M. Hon et al. (2021), which have a mean fractional uncertainty of 11.4%, to those from J. T. Mackereth et al. (2021). For clarity in the comparison, we exclude stars with $\nu_{\rm max} < 10\ \mu$Hz or differences exceeding 20% (but retain them in the overall sample), finding a mean fractional offset of 2.9% between the two catalogs, demonstrating strong overall consistency (see Figure 1).

These studies together indicate that the M. Hon et al. (2021) $\nu_{\rm max}$ values are reliable, and therefore we follow D. Stello et al. (2008) and M. Hon et al. (2021) to estimate masses using the following equation:

$$\frac{M}{M_{\odot}} = \left(\frac{\nu_{\rm max}}{\nu_{\rm max,\odot}}\right)\left(\frac{R}{R_{\odot}}\right)^{2}\left(\frac{T_{\rm eff}}{T_{\rm eff,\odot}}\right)^{0.5},$$

with $\nu_{\rm max,\odot} = 3090\,\mu$Hz (D. Huber et al. 2011) and $T_{\rm eff,\odot} = 5771.8$ K (A. Prša et al. 2016). We omit the $f_{\nu_{\rm max}}$ correction proposed by M. H. Pinsonneault et al. (2025) because such corrections are pipeline specific and may overcompensate for systematics in $\Delta\nu$ scaling relations (A. L. Ash et al. 2025). Tests on a subset of our sample show that including $f_{\nu_{\rm max}}$ alters masses by <3%, which is within our quoted uncertainties.

While the agreement is satisfactory, caveats exist for the method of $\nu_{\rm max}$ determination. Blending remains a potential issue, particularly for stars with similar $\nu_{\rm max}$ values (M. Hon et al. 2021). Fainter targets are at the greatest risk of blending. The remaining offset between the $\nu_{\rm maxes}$ in M. Hon et al. (2021) and those in J. T. Mackereth et al. (2021) may also be related to the treatment of convective backgrounds, and future work may recommend additional calibration (M. H. Pinsonneault et al. 2018; S. Kalarickal Ramachandra et al. 2024).

### *2.2. Gaia Radii*

Because we do not have seismic $\Delta\nu$ values, we require an additional constraint to estimate masses. We use radii from M. Hon et al. (2021) computed using `isoclassify` (D. Huber et al. 2017; T. A. Berger et al. 2020, 2023), which uses parallaxes from Gaia EDR3 (Gaia Collaboration et al. 2021) as an input. The offsets between the two temperature scales are on the order of a few percent, and we conservatively include error floors to account for potential systematics (see Section 6). To eliminate blended targets and specify isolated ones, they required that each target have an effective temperature ($T_{\rm eff}$) and radius ($R$) that are similar to those of an oscillating red giant. In this procedure (M. Hon et al. 2021), extinction corrections were applied to Two Micron All Sky Survey $K$-band magnitudes to obtain apparent magnitudes. $A_V$ values were computed for each star using combined three-dimensional dust maps (R. Drimmel et al. 2003; D. J. Marshall et al. 2006; G. M. Green et al. 2019), as implemented in the `mwdust` package (J. Bovy et al. 2016), assuming an extinction uncertainty of 0.02 mag. Bolometric corrections were then derived by interpolating $T_{\rm eff}$, $\log g$, [Fe/H], and $A_V$ within the Mesa Isochrones and Stellar Tracks (MIST)/C3K bolometric correction tables (J. Choi et al. 2016), adopting $M_{\rm bol,\odot} = 4.74$ mag to reproduce the solar luminosity. These bolometric corrections (uncertainty ≈ 0.02 mag) were applied to the extinction-corrected apparent magnitudes to compute absolute magnitudes and stellar luminosities.

### *2.3. Spectroscopic Sample and Calibration*

#### *2.3.1. XGBoost*

The Gaia mission (Gaia Collaboration et al. 2016, 2018), renowned for its precise astrometry and photometric measurements, also provides spectroscopic data through its Radial Velocity Spectrometer (RVS) and low-resolution XP spectra. The short-wavelength spectra are referred to as BP spectra, while the long-wavelength spectra are known as RP spectra. XP spectra are used to collectively refer to both types of Gaia spectra. The RVS operates at a resolution of ∼8000 around the near-infrared Ca triplet, enabling the measurement of radial velocities and contributing to our understanding of Galactic dynamics. Additionally, XP spectra, obtained via the BP and RP prisms, span wavelengths from 350 to 1000 nm but have limited resolution ($R \sim 40$–150), posing challenges for deriving precise metallicities ([M/H]). To address these limitations, a data-driven approach using the machine learning algorithm XGBoost was employed (R. Andrae et al. 2023) to estimate stellar parameters such as [M/H], effective temperature ($T_{\rm eff}$), and surface gravity ($\log g$). The XGBoost model was trained on high-fidelity datasets, including Apache Point Observatory Galactic Evolution Experiment (APOGEE) DR17 (see Section 2.4), which provides stellar parameters for stars with well-characterized metallicities. Training features incorporated XP spectral coefficients, Gaia XP-derived synthetic photometry (e.g., Gaia Collaboration 2022), broadband magnitudes, and parallaxes, effectively mitigating degeneracies, such as those between $T_{\rm eff}$ and dust reddening. By leveraging these features and augmenting the training set with metal-poor stars, XGBoost delivers robust metallicity estimates across a wide range of stellar types and magnitudes, demonstrating its potential as a powerful tool for deriving metallicities in stellar populations.

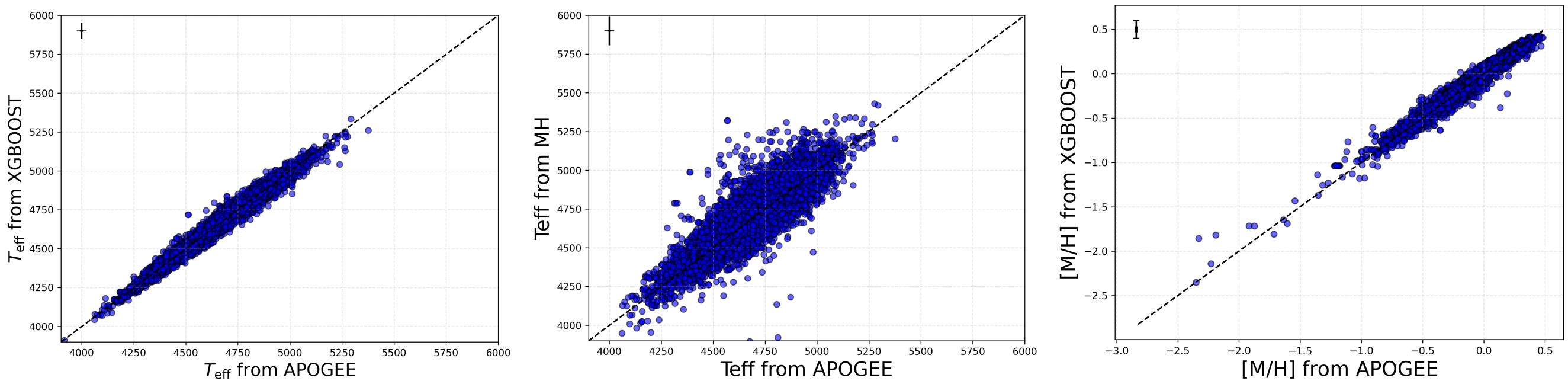

**Figure 2.** Comparisons between the parameters inferred from APOGEE, and those inferred from Gaia. In particular, we find that the XGBoost temperature scale (left) is closer to the APOGEE temperature scale than the temperatures computed in M. Hon et al. (2021, middle). We also find that the XGBoost metallicities agree with the APOGEE scale to better than 0.03 dex in the majority of cases (right) thus validating their use in our age calculations. The cross located in the top left corner of each plot represents the average uncertainty across each axis, with the ($T_{\rm eff}$) error from XGBoost being uniformly 50 K.

We employ a crossmatch between the Gaia Data Release 3 (DR3) dataset and our TESS-identified red giants, creating a catalog of 134,901 stars that includes both TESS seismology and Gaia-derived radii and metallicities. This crossmatch used the full Gaia DR3 sample from R. Andrae et al. (2023), rather than the vetted subset in their Table 2. While R. Andrae et al. (2023) report typical [M/H] uncertainties of ∼0.08–0.10 dex depending on color, they do not provide full propagated errors due to the machine learning approach; instead, we estimate uncertainties empirically from our APOGEE comparison (see Figure 2), finding that our XGBoost metallicities are consistent to within ∼0.001 dex on average, with a scatter of 0.04 dex, which we adopt and propagate throughout the analysis.

### 2.4. APOGEE

For further calibration and validation of the XGBoost parameters, we employ spectroscopic parameters from the APOGEE DR17 (Abdurro'uf et al. 2022) run as part of the third generation Sloan Digital Sky Survey (SDSS)/APOGEE and fourth-generation SDSS/APOGEE-2 surveys (S. R. Majewski et al. 2017). These spectra offer estimates of parameters such as effective temperature, metallicity, and surface gravity that are more precise and accurate than those derived from the Gaia data, which provide reduced uncertainties in our derived masses and ages. We can use the subset of stars with available APOGEE data to validate and calibrate our mass and age estimates, along with their associated uncertainties.

The APOGEE Stellar Parameters and Chemical Abundances Pipeline (A. E. García Pérez et al. 2016; J. A. Holtzman et al. 2018; H. Jönsson et al. 2020) processes the spectroscopic data using the SYNSPEC atmospheric models (I. Hubeny & T. Lanz 2017; I. Hubeny et al. 2021). This automated pipeline provides near-infrared spectra and derives stellar parameters such as effective temperature ($T_{\rm eff}$), surface gravity ($\log g$), and metallicity ([M/H]).

Comparisons with clusters (A. Sinha et al. 2024) and external validation data (M. H. Pinsonneault et al. 2025) suggest that the uncertainties on the abundances and stellar parameters are small, although there may be some residual trends at the 0.03 dex level with temperature and gravity (A. Sinha et al. 2024).

Because these data have been extensively processed and validated, many low-level trends and offsets have been documented. Potentially relevant to our study, we note that the APOGEE gravities for clump stars show a slight correlation between ($T_{\rm eff}$) and $\log g$ at the 0.18 dex per 1000 K level (Y.-Q. Chen et al. 2015; J. A. Holtzman et al. 2015), and APOGEE gravities for stars with $[{\rm M/H}] < -1.5$ dex may be offset by 0.2 dex from optical results (H. Jönsson et al. 2018). These offsets could affect our inferred ages, which rely on metallicity. We find that at low metallicity, two stars with the same mass and $\log g$ but a metallicity difference of 0.2 dex have an age difference of $1.11 \pm 0.82$ Gyr. Similarly, after refining the sample to include only RGB stars, the sensitivity of stellar age to surface gravity was found to be $\frac{d\,{\rm Age}}{d\,\log g} \approx 0.03$ Gyr dex$^{-1}$, corresponding to an age change of $\Delta{\rm Age} \approx 0.01 \pm 0.02$ Gyr for a difference of $\Delta \log g = 0.18$ dex. This result indicates that age estimates are largely insensitive to small uncertainties in $\log g$ for this sample.

### 2.5. Entire Sample

For our sample, we follow the procedure of A. T. Theodoridis & J. Tayar (2023) to combine Gaia, TESS, and XGBoost data to estimate seismically informed masses and surface gravities for each of our red giants.

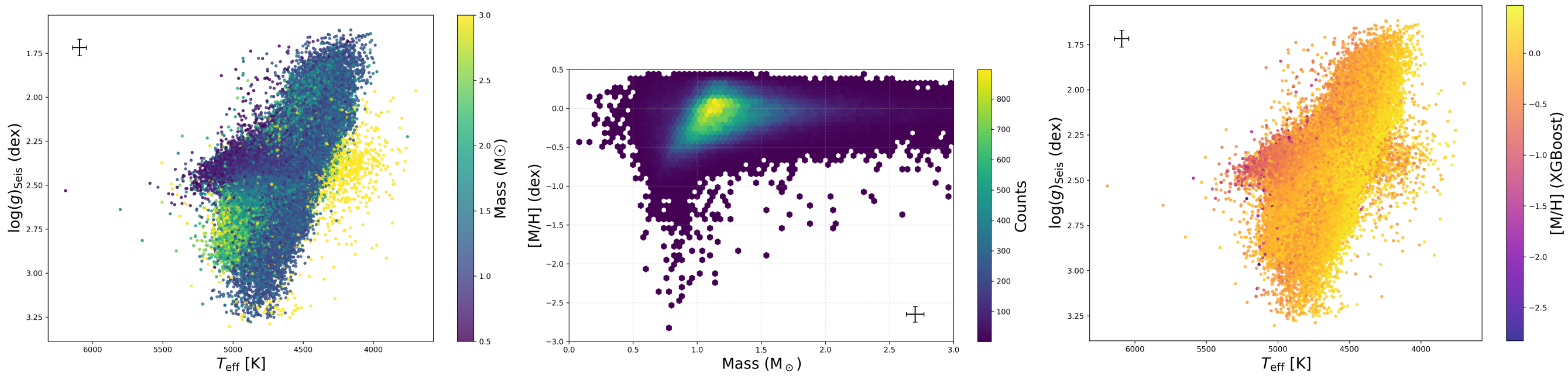

**Figure 3.** We present two H-R diagrams of our sample, showing the relationship between seismic surface gravity ($\log(g)_{\rm Seis}$) and XGBoost effective temperature ($T_{\rm eff}$), color coded by metallicity (right) and mass (left). The middle plot shows the density distribution of the sample, highlighting the range of stellar masses and metallicities. We also note that the population of very cool, apparently high-mass ($>3\ M_\odot$) yellow stars in the left panel are likely to be spurious, and not genuine rapidly rotating or spotted massive merger products. In general, we observe that giants near solar metallicity span a range of masses, while lower-metallicity stars tend to have lower mass, and no giants exceed a metallicity of approximately [M/H] $\sim$ 0.5. These plots provide a detailed overview of our sample's distribution across these key parameters. The cross in the corner of each plot represents the average uncertainty across each axis.

The crossmatch occurs through a Python script utilizing the `astropy` table function, ensuring proper correspondence of stellar IDs between the applicable datasets. Specifically, we used the correspondence between `TIC_ID` from the TESS Input Catalog (v8; K. G. Stassun et al. 2019) and `gaiadr3_source_id` and eliminated any stars that did not have a matching counterpart across the individual datasets. Once this crossmatch was completed, we were left with 134,901 red giant stars out of the 158,014 red giant oscillators detected in our TESS sample.

We added to our table calculations of seismic $\log g$ and mass in a similar fashion to M. Hon et al. (2021), using Gaia EDR3 radius (M. Hon et al. 2021), XGBoost ($T_{\rm eff}$), and $\nu_{\rm max}$ (M. Hon et al. 2021) data to combine these different datasets and estimate seismically informed masses and surface gravities for all 134,901 stars in our sample. We used the APOGEE DR17 sample as an additional check to validate our choice of effective temperature and [M/H] parameters. For the $\sim$12,000 stars where APOGEE parameters were available, we compared the ($T_{\rm eff}$) and [M/H] values with those from XGBoost (Figure 2) versus M. Hon et al. (2021). Figure 3 shows the H-R diagram positions, as well as the mass and metallicity distributions, for the full sample.

Unfortunately, our current sample does not overlap with the sample of wide binaries (K. El-Badry et al. 2021) or stellar clusters (T. Cantat-Gaudin et al. 2020), which would allow us to perform a empirical check on our uncertainty assumptions (e.g., Y. Li et al. 2025). We expect this to be likely due to challenges of processing data from TESS's large pixels (21$''$ in size) in crowded fields, although we encourage future work on these important samples.

The determination of error bars for the seismic $\log g$ and mass values was calculated using standard equations for error propagation, and the average errors are 12% for mass and 5% for seismic $\log g$.

### *2.6. Data Validation*

In general, the frequency of maximum power, $\nu_{\rm max}$, is expected to scale with the surface gravity and temperature of the star. We also note that the temperature values from XGBoost generally agree with the APOGEE temperatures better than the temperature values inferred using `isoclassify` in M. Hon et al. (2021; see Figure 2), and therefore we use the XGBoost temperatures in our calculations.

## 3. Modeling Grid

Our modeling framework employs Modules for Experiments in Stellar Astrophysics (MESA), version r22.11.1 (B. Paxton et al. 2011, 2013, 2015, 2019; A. S. Jermyn et al. 2023), to construct a comprehensive grid of stellar models tailored to the properties of low-mass red giants. MESA's adaptive mesh refinement and advanced time-stepping algorithms allow it to handle a broad range of stellar evolution scenarios (binary systems, binary mergers, mass loss, mass transfer, rotation, etc.), making it an ideal choice for simulating red giant evolution across diverse metallicities and other stellar parameters such as age, mass, and $\log g$. We note that our stellar evolution models do not include nonstandard processes such as rotation, thermohaline mixing, rotational mixing, or binary interactions. As demonstrated in previous studies (Y. Lebreton et al. 2014), these effects typically introduce age variations of $\lesssim$10% for RGB stars at fixed mass. To account for this systematic uncertainty, we have incorporated a conservative error floor in our age estimates (see Section 6). While these processes could be explored in future work, their exclusion here is consistent with standard analysis approaches for large samples such as ours.

The MESA models are particularly robust in calculating evolutionary tracks for low-mass stars, from the zero-age main sequence (ZAMS) through to the asymptotic giant branch (AGB) phase. In particular, MESA solves the one-dimensional stellar-structure and composition equations for each star we model. We use MESA as it is capable of running through the helium flash consistently, which is needed for our grid and age interpolation. MESA can successfully evolve models through the helium flash without requiring special user-implemented treatments when appropriate numerical settings are used (time step controls and spatial resolution parameters as described in Appendix). While MESA internally adjusts its convergence criteria during the flash (e.g., temporarily relaxing luminosity constraints; B. Paxton et al. 2013), these are automatic algorithmic adaptations that do not require manual intervention. Processes for mixing and burning are solved simultaneously, and discussion of specific stellar input physics are described below.

We have constructed our own modeling grid as follows:

$$M_{\rm init} = \{0.7,\ 0.8, \ldots, 3.0\},$$
$$[{\rm Fe/H}]_{\rm init} = \{-2.0,\ -1.9, \ldots, +0.5\},$$

**Table 1**
Summary of the Input Physics Used in Our Model Grids Compared to M. Joyce et al. (2024)

| Parameter | Grid | Joyce |
| --- | --- | --- |
| Atmosphere | A. S. Eddington (1930) | A. S. Eddington (1930) |
| $\alpha$ Enhancement | None | None |
| Convective Overshoot | diffusive $f_{\rm core} = 0.016$, $f_{\rm env} = 0.0174$ (J. Choi et al. 2016) | None |
| Diffusion | None | None |
| Equation of State | OPAL + SCVH + MacDonald + HELM + PC | OPAL + SCVH + MacDonald + HELM + PC |
| High-temperature Opacities | OPAL (F. J. Rogers et al. 1996) | OPAL (C. A. Iglesias & F. J. Rogers 1996) |
| Low-temperature Opacities | J. W. Ferguson et al. (2005) | ÆSOPUS (P. Marigo & B. Aringer 2009) |
| Mixing Length | 1.94 | 1.931 |
| Mixture | N. Grevesse & A. J. Sauval (1998) | N. Grevesse & A. J. Sauval (1998) |
| Nuclear Reaction Rates | R. Kunz et al. (2002) | B. Paxton et al. (2018) |
| Weak Screening | A. I. Chugunov et al. (2007) | None |
| Scaling Factor | D. Reimers (1975) $\eta = 0.2$, T. Bloecker (1995) $\eta = 0.01$ | T. Bloecker (1995) $\eta = 0.01$ |
| Solar $X$ | 0.7094815 | 0.7491 |
| Solar $Y$ | 0.2725693 | 0.2377 |
| Solar $Z$ | 0.0179492 | 0.0133 |
| [M/H] Range | −2.0 to +0.5 | −1.2 to +1.0 |
| Mass Range | 0.7–3.0 $M_\odot$ | 1.0 to 5.0 $M_\odot$ |

$$[{\rm Fe/H}] = \log_{10}\left(\frac{Z/X}{(Z/X)_\odot}\right),$$

$$Z = Z_\odot \times 10^{[{\rm Fe/H}]},$$

which corresponds to

$$Z_{\rm init} \in \{0.000179492, 0.000225967040014, 0.000284475648901, 0.000358133623438, 0.000450863519365, 0.000567603541779, 0.00071457052257, 0.000899590989382, 0.00113251795675, 0.00142575563507, 0.00179492, 0.00225967040014, 0.00284475648901, 0.00358133623438, 0.00450863519365, 0.00567603541779, 0.0071457052257, 0.00899590989382, 0.0113251795675, 0.0142575563507, 0.0179492, 0.0225967040014, 0.0284475648901, 0.0358133623438, 0.0450863519365, 0.0567603541779\},$$

for a total of $24 \times 26 = 624$ models in our grid. Our $Y_i$ is scaled according to $Z_{\rm init}$ via the canonical relation

$$Y_i = Y_0 + \frac{\Delta Y}{\Delta Z} \times Z_i, \quad (1)$$

where $Y_0$ is the primordial He abundance and $\frac{\Delta Y}{\Delta Z}$ is the He-to-metal enrichment ratio. We take $Y_0 = 0.2454$ (Planck Collaboration et al. 2020) and $\frac{\Delta Y}{\Delta Z} = 1.5137$ (see Section 3.2).

### 3.1. Adopted Stellar Physics

We used a simplified version of the input physics described in M. Joyce et al. (2024) to construct a grid with updated physics specifically tailored to the mass and metallicity ranges present in our sample, focusing on the factors most likely to significantly impact the inferred ages of red giants. The metallicity range in our grid is set based on the distribution we found in our completed crossmatched dataset between Gaia, APOGEE, and TESS as detailed in the previous sections. The percentage of models that reached the RGB tip of their evolution is roughly 93%. With computational efficiency in mind, we imposed a 24 hr runtime limit per model. This allowed all models to fully traverse the RGB, with many successfully completing core-helium burning within this time frame. We explicitly aim for models to reach the AGB for completeness, but consider models that run at least to the helium flash as "sufficient." Of all the models run in the original grid, there was a systematic error in the [Fe/H] value of 0.3, where the model failed across all masses. The discontinuity likely occurs because MESA switches internal tables near this [Fe/H] value, creating numerical convergence challenges. While we cannot definitively identify the root cause without deeper investigation of MESA's internals, we have documented the effect empirically to alert users of this boundary condition. Future work could systematically probe this metallicity transition. In our finalized grid, we implemented a value of 0.299 instead, and had greater success with only five models failing. With the incomplete tracks excluded, our grid consists of 619 models in total.

We show in Table 1 a synopsis of the input physics included in our models, and include a sample inlist `inlist_common` file in Appendix. In particular, we assume an N. Grevesse & A. J. Sauval (1998) solar mixture, the Eddington gray $T$–$\tau$ relation for atmospheric boundary conditions (A. S. Eddington 1930), and the Henyey mixing length scheme (L. G. Henyey et al. 1964) with a fixed value of $\alpha_{\rm MLT} = 1.94$ based on solar calibration in our grid (see Section 3.2). For high-temperature opacities we used the OPAL (F. J. Rogers et al. 1996) tables with the appropriate GS98 mixture, and for low temperatures we used the J. W. Ferguson et al. (2005) opacities with the appropriate GS98 mixture in order to produce more accurate predictions for low-temperature giants. The models utilize the `pp_extras.net`, `co_burn.net`, and `approx21.net` nuclear reaction networks. In particular, the `approx21` nuclear reaction network consists of 21 species: $^{1}$H, $^{3}$He, $^{4}$He, $^{12}$C, $^{14}$N, $^{16}$O, $^{20}$Ne, $^{24}$Mg, $^{28}$Si, $^{32}$S, $^{36}$Ar, $^{40}$Ca, $^{44}$Ti, $^{48}$Cr, $^{52}$Fe, $^{54}$Fe, $^{56}$Fe, $^{56}$Ni, protons, and neutrons. We use the Reimers and Blöcker schemes for cool RGB and AGB winds, respectively (D. Reimers 1975; T. Bloecker 1995). We employ a value of $\eta_{\rm Reimers} = 0.2$ (A. Miglio et al. 2012) for the Reimers RGB wind scheme. For the Blöcker wind scheme

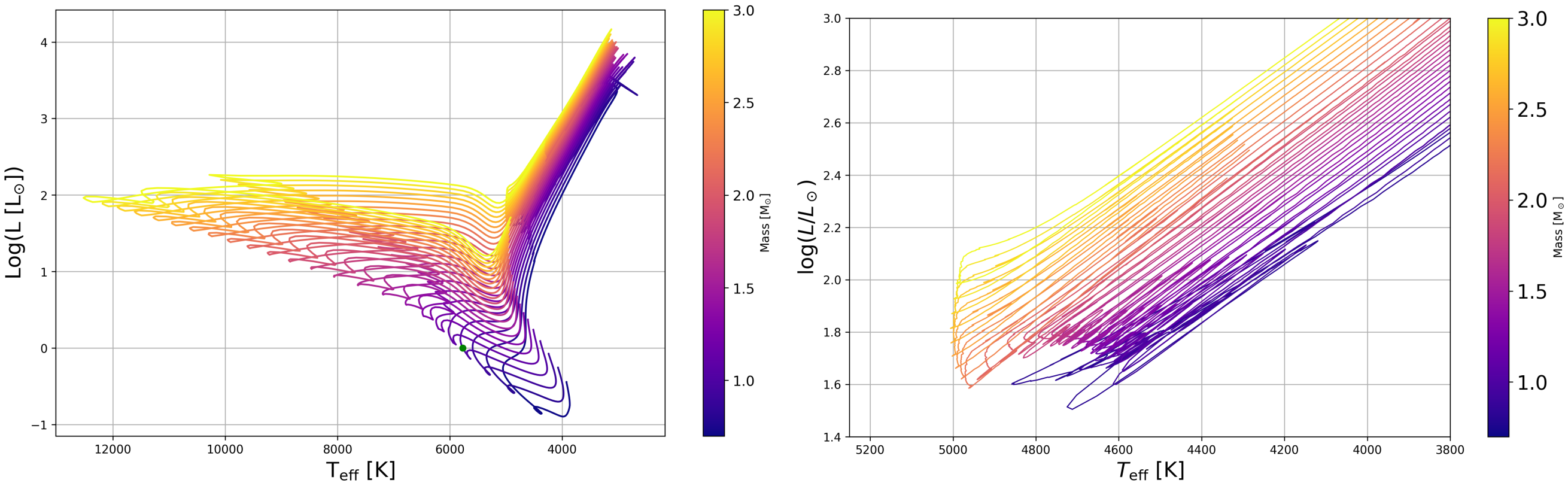

**Figure 4.** Stellar evolutionary tracks for various masses at solar metallicity in our grid. The left panel shows the full tracks from the pre–main sequence through the ZAMS, the RGB, and up to the clump. The right panel shows a zoomed-in view of the post-helium-flash, core He-burning phase (red clump (RC) and horizontal branch), where the morphology depends sensitively on the adopted stellar physics (see L. Girardi 2016). The green dot indicates the parameters of the Sun, and the solar-mass track passes through this point at the appropriate age.

(T. Bloecker 1995), we adopted a value of $\eta_{\text{Blöcker}} = 0.01$, similar to J. Choi et al. (2016, who used MIST). We use the Schwarzschild criterion for convective stability (F. Anders et al. 2023) and since we use the Schwarzschild criterion, we do not account for thermohaline mixing or semiconvection. Rotation is neglected because it does not significantly affect age interpolation results for red giants. Although previous studies have noted age differences upward of 20% after implementation of rotation (Y. Lebreton et al. 2014) for higher-mass stars (e.g., 40 $M_\odot$), 75% of the stars in our sample fall below the the Kraft break (A. C. Beyer & R. J. White 2024) on the main sequence and 96% are less than 2.1 $M_\odot$, which would correspond to fast rotation rates of 154 km s$^{-1}$ on average (J. Zorec & F. Royer 2012). As shown by J. Zorec & F. Royer (2012), the rotational evolution of stars below $\sim$2.5 $M_\odot$ leads to minor changes in their main-sequence lifetimes, with rotation altering inferred ages by at most $\sim$10% of the total evolutionary timescale, so we assume rotation is not a significant contributor to the age uncertainty. Furthermore, we focus solely on single-star evolutionary calculations.

We describe how these mass-loss choices compare to empirically calibrated mass-loss rates in Section 4, and describe our final mass-loss assumptions for clump stars there as well.

### 3.2. Modeling Procedure

To perform a solar calibration, we varied the mixing length and helium abundances until the solar-mass (1.00 $M_\odot$) and solar-metallicity ($z = 0.0179492$) model matched the solar temperature (5772 K; A. Prša et al. 2016) and solar luminosity ($L_{\text{Sun}} = 3.8270 \times 10^{33}$ erg s$^{-1}$ ) at the solar age (4.57 Gyr; J. N. Bahcall et al. 2006; see Table 1). We adopt initial bulk metallicity ($Z$) for simplicity, as RGB mixing largely erases surface composition variations from main-sequence diffusion. While the mixing length and helium abundance may vary with metallicity and evolutionary state (J. Tayar et al. 2017; Y. Li et al. 2024), studies demonstrate that such variations introduce minimal bias in red giant properties when compared to asteroseismic constraints (J. Tayar et al. 2022; L. Morales et al. 2025). We account for residual systematic uncertainties in our error budget (see Section 6) and present our tracks for a range of masses at solar metallicity in Figure 4 for reference.

## 4. Age Interpolation

Our primary goal is to estimate the ages of stars to further our understanding of stellar evolution. To achieve this, we use `kiauhoku`,[2] an open-source package designed to interpolate stellar parameters from evolutionary model tracks (Z. R. Claytor et al. 2020). `kiauhoku` enables custom grid installation, where we resample the model tracks into equivalent evolutionary points (EEPs; A. Dotter 2016). On the RGB, these are usually assigned using core mass. Stellar parameters are then interpolated as a function of initial mass, metallicity, and EEP (J. Tayar et al. 2022). We use the `gridsearch_fit` routine within `kiauhoku`, which performs a purely deterministic grid-based lookup and evaluates a mean, providing the initial guess for a Nelder–Mead optimization (J. A. Nelder & R. Mead 1965) that minimizes a mean squared error loss function (L. Morales et al. 2025). The optimization is considered successful when it converges to a solution within a tolerance of $10^{-6}$. The spacing between mass and age is nonlinear, meaning that weights calculated using a two-point (linear) interpolation will not adequately capture this relationship, as it depends only on the distance from the lower-mass track. Therefore, we use a four-point (cubic) interpolation scheme where possible, to capture this nonlinear behavior in age. However, in regions where many evolutionary tracks fail to reach the tip of the giant branch, typically at low masses and high metallicities (i.e., old ages), we simplify to a two-point interpolation scheme to provide an estimate of age in as many stars as possible. We note that the heterogeneous spacing in age among tracks can introduce a "striping" effect (see Figure 9), caused by the interpolation weight being biased toward the lower-mass track. We use the `UnivariateSpline` function from the `scipy.interpolate` package, which fits a 1D spline through the selected data points. Specifically, we interpolate our model grid in mass, [Fe/H], and EEP space to evaluate stellar properties at the observed parameters: (1) mass derived from $\nu_{\max}$, Gaia $R$, and XGBoost $T_{\text{eff}}$; (2) surface gravity ($\log g$) computed from $\nu_{\max}$ and XGBoost $T_{\text{eff}}$; and (3) XGBoost metallicity. This standard approach mirrors interpolation methods widely used in stellar evolution studies (e.g., B. Paxton et al. 2013; A. Dotter 2016). For clump stars, we adjust their masses to account for mass

[2] https://github.com/zclaytor/kiauhoku

**Table 2**
Our Collected Data and Computed Values including Spectroscopic, Gaia-derived, and Asteroseismic Results, as Well as Derived Masses and Ages (See Text for Details)

| Label | Units | Contents |
|---|---|---|
| TIC | None | Stellar ID in TESS |
| Star_type | None | Recommended evolutionary state |
| Final_age | Gyr | Age based on evolutionary state |
| $\nu_{\max}$ | $\mu$Hz | Frequency of maximum power (M. Hon et al. 2021) |
| Radius_gaia | $R_\odot$ | Radius from Gaia (M. Hon et al. 2021) |
| Teff_xgboost | K | $T_{\rm eff}$ from Gaia XGBoost |
| M_H_xgboost | dex | [Fe/H] metallicity from Gaia XGBoost |
| Logg_xgboost | cgs | Spectroscopic log$g$ from XGBoost |
| Logg_seis | cgs | Seismic log$g$ |
| E_Logg_seis | cgs | Uncertainty on seismic log$g$ |
| Mass_seis | $M_\odot$ | Seismic mass |
| E_Mass_seis | $M_\odot$ | Uncertainty on seismic mass |
| Initial_mass | $M_\odot$ | Assumed initial mass of star accounting for empirical mass loss |
| Teff_rgb | K | Model temperature assuming RGB |
| Teff_rc | K | Model temperature assuming RC |
| Median_age_rgb | Gyr | Median age assuming RGB |
| Median_age_rc | Gyr | Median age assuming RC |
| E_lower_age_rgb | Gyr | Lower confidence interval RGB age |
| E_upper_age_rgb | Gyr | Upper confidence interval RGB age |
| E_lower_age_rc | Gyr | Lower confidence interval RC age |
| E_upper_age_rc | Gyr | Upper confidence interval RC age |
| Teff_diff | K | Difference in effective temperature (xgboost – rgb) |
| Flag | None | Marked 0 if model fit has converged |

**Note.** This table is available in its entirety on Zenodo at doi:10.5281/zenodo.18050974.

(This table is available in its entirety in machine-readable form in the online article.)

loss, as described in the following section. Uncertainties in the age estimates are calculated by propagating the observational errors through a Monte Carlo simulation of 100 iterations, with each iteration performing the same deterministic grid-based interpolation process. Finally, we present the resulting age distribution in Section 6, which shows the relationship between age and metallicity for both clump stars and first ascent giants.

We do not compute ages for stars with masses below 0.7 $M_\odot$ even though such stars represent 1.2% of our sample (1623 stars), as single-star evolution would imply unphysical ages exceeding the cosmic timescale. We expect that some of them are binary interaction products and some may be related to measurement errors, but we consider detailed analysis of their properties outside the scope of this study. To emphasize stars with physically plausible ages, we apply a 14 Gyr upper limit in all figures; the complete sample (including older and/or unphysical ages) is provided in Table 2. We also do not compute ages for stars above 3 $M_\odot$, which make up 1696 stars (or 1.26%) of our sample, as our initial investigation into the temperatures, evolutionary states, and log $g$ values suggest that these stars are not preferentially located where we would expect massive stars; we therefore infer that most of them are listed as massive from measurement uncertainties rather than being truly massive (C. L. Crawford et al. 2025).

For each star, we calculate ages under three evolutionary scenarios: (1) RGB assumption (no mass loss): current

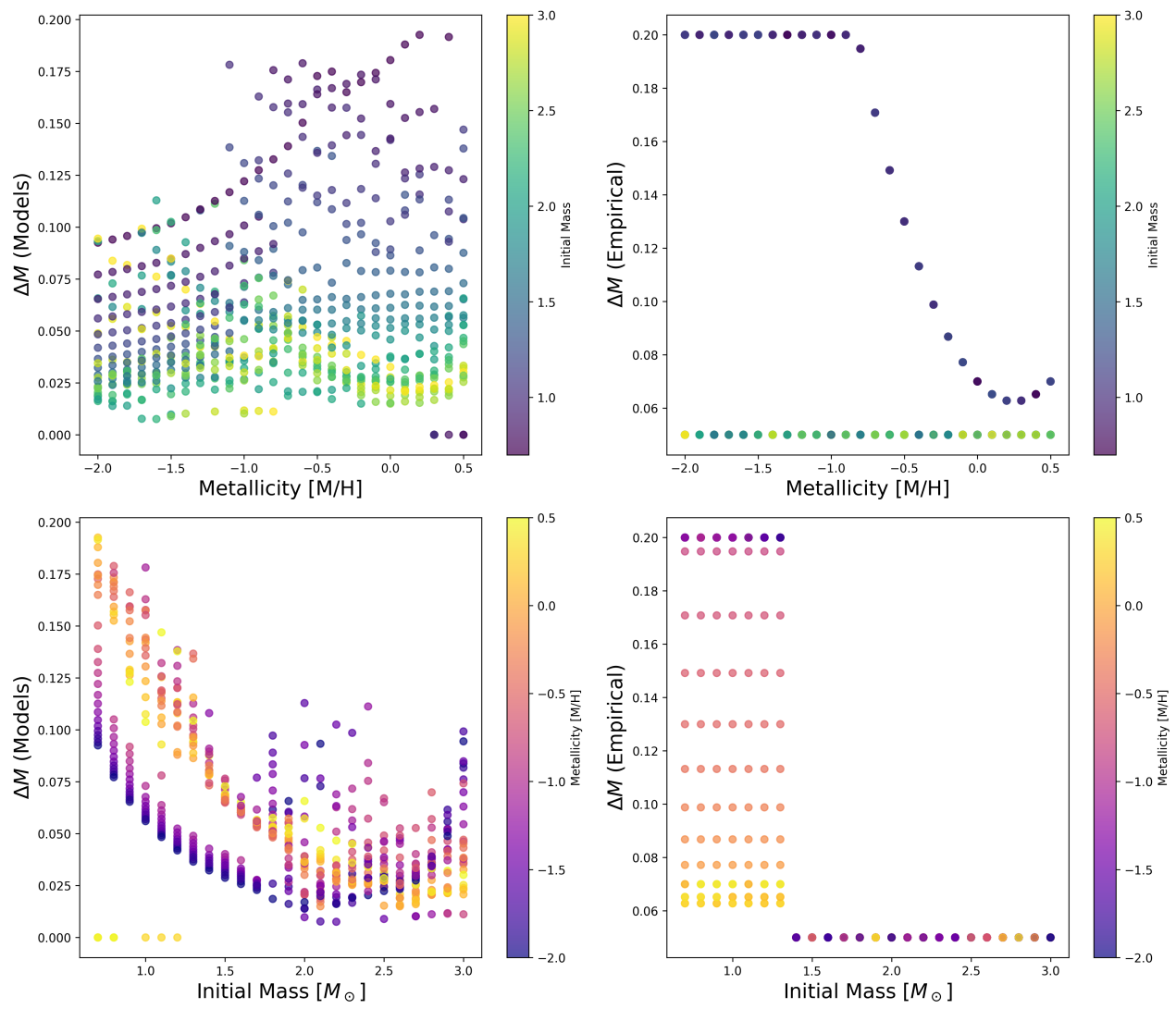

**Figure 5.** Predicted mass loss as a function of stellar mass and metallicity. The figure compares two mass-loss schemes: one based on theoretical model tracks (left) and the other using empirical mass-loss calibrations from J. Tayar et al. (2023) and M. H. Pinsonneault et al. (2025, both on right). While theoretical predictions have a strong mass dependence and a weak metallicity dependence, the current empirical mapping is the opposite. We expect real mass-loss rates to depend on both mass and metallicity. Therefore we compute ages with both sets of assumptions.

mass = birth mass (Option 1); (2) empirical clump assumption: adopting our observational mass-loss prescription (Option 2); and (3) theoretical clump assumption: using model-based mass loss (Option 3, see Figure 5).

We present our final converged ages and determined evolutionary states in Table 2, along with an alternative list of available variables in Table 3. This comprehensive dataset is structured to allow future users the flexibility to reclassify evolutionary states or apply alternative classification schemes such as probabilistic assignment based on their own goals. Variables ending in "rgb" correspond to values calculated under the assumption that the star is on the RGB, while those ending in "rc" reflect calculations assuming the star is in the RC. The final adopted evolutionary state for each star is indicated in the "Star_Type" column; however, both sets of variables are included in the dataset to allow for flexible reanalysis and completeness.

## 5. Mass Loss

Mass loss occurs before stars reach the RC, which means that the current masses of clump stars are not their birth masses. If this mass loss is not properly accounted for, it can introduce significant errors in age estimates.

### *5.1. Mass-loss Prescriptions*

There are three general approaches to estimating stellar mass loss: one assumes that all stars in the sample are on the RGB and therefore neglects mass loss, another derives it directly from stellar evolution models, and the third relies on empirical calibrations. The models are calculated using prescriptions of mass loss that scale with the atmospheric properties (e.g., D. Reimers 1975), while empirical calibrations are based on observed data from different stars in the same environment.

**Table 3**
Our Collected Data and Computed Values Including Spectroscopic, Gaia-derived, and Asteroseismic Results, as Well as Derived Masses and Ages (See Text for Details) Assuming Our Alternate, Model-based Mass-loss Scheme for Red Clump Stars

| Label | Units | Contents |
| --- | --- | --- |
| TIC | None | Stellar ID in TESS |
| Star_type | None | Recommended evolutionary state |
| Final_age | Gyr | Age based on evolutionary state |
| $\nu_{max}$ | $\mu$Hz | Frequency of maximum power (M. Hon et al. 2021) |
| Radius_gaia | $R_\odot$ | Radius from Gaia (M. Hon et al. 2021) |
| Teff_xgboost | K | $T_{eff}$ from Gaia XGBoost |
| M_H_xgboost | dex | [Fe/H] metallicity from Gaia XGBoost |
| Logg_xgboost | cgs | Spectroscopic log$g$ from XGBoost |
| Logg_seis | cgs | Seismic log$g$ |
| E_Logg_seis | cgs | Uncertainty on seismic log$g$ |
| Mass_seis | $M_\odot$ | Seismic mass |
| E_Mass_seis | $M_\odot$ | Uncertainty on seismic mass |
| Initial_mass | $M_\odot$ | Assumed initial mass of star accounting for model-derived mass loss |
| Teff_rgb | K | Model temperature assuming RGB |
| Teff_rc | K | Model temperature assuming RC |
| Median_age_rgb | Gyr | Median age assuming RGB |
| Median_age_rc | Gyr | Median age assuming RC |
| E_lower_age_rgb | Gyr | Lower confidence interval RGB age |
| E_upper_age_rgb | Gyr | Upper confidence interval RGB age |
| E_lower_age_rc | Gyr | Lower confidence interval RC age |
| E_upper_age_rc | Gyr | Upper confidence interval RC age |
| Teff_diff | K | Difference in Effective Temperature ($T_{eff}$ from Gaia xgboost – rgb) |
| Flag | None | Marked 0 if model fit is converged |

**Note.** This table is available in its entirety on Zenodo at doi:10.5281/zenodo.18050974.

(This table is available in its entirety in machine-readable form in the online article.)

For our analysis, we explored three options for mass loss: (1) assuming all stars are RGB without mass loss; (2) model based, which calculates mass loss based on the theoretical model tracks in our grid ($\eta_{\rm Reimers} = 0.2$; A. Miglio et al. 2012); and (3) empirically calibrated mass loss, which uses the difference in asteroseismic mass between $\alpha$-enhanced RGB and clump stars to estimate a metallicity dependence to the mass loss, as described in the alternative age inference method from M. H. Pinsonneault et al. (2025; see also J. Tayar et al. 2023; K. Brogaard et al. 2024; Y. Li et al. 2025; C. Marasco et al. 2025). We present the results from Options 1, 2, and 3 within our Tables 2 and 3, along with the predicted mass loss from these schemes as a function of mass and metallicity in Figure 5.

Both methods have limitations. The empirical mass-loss prescription is applied across our full metallicity range ([Fe/H] = −2 to +0.5; see Figure 5), though extrapolation beyond this calibrated parameter space may overestimate mass loss at low metallicity and high mass. While globular clusters provide constraints at low [Fe/H] (e.g., M. Tailo et al. 2020; M. Howell et al. 2025), our empirical trends remain less certain in this regime. On the other hand, the theoretical mass loss from the tracks exhibits a different metallicity trend that is inconsistent with observational data (M. Howell et al. 2022; M. Tailo et al. 2022; C. Marasco et al. 2025; M. H. Pinsonneault et al. 2025). For our main table (Table 2), we decided to use the empirically calibrated mass loss to better capture the metallicity dependence, but we added an additional mass dependence calibrated on the models within our grid to better account for the expected and observed mass dependence of RGB mass loss. For our empirical mass loss, we used the following equation from M. H. Pinsonneault et al. (2025) and applied it to all stars within our dataset:

$$\Delta{\rm mass} = (0.12 \cdot ([{\rm M/H}] + 1.0)^2 + 0.95) - (0.3 \cdot [{\rm M/H}] + 1.0).$$

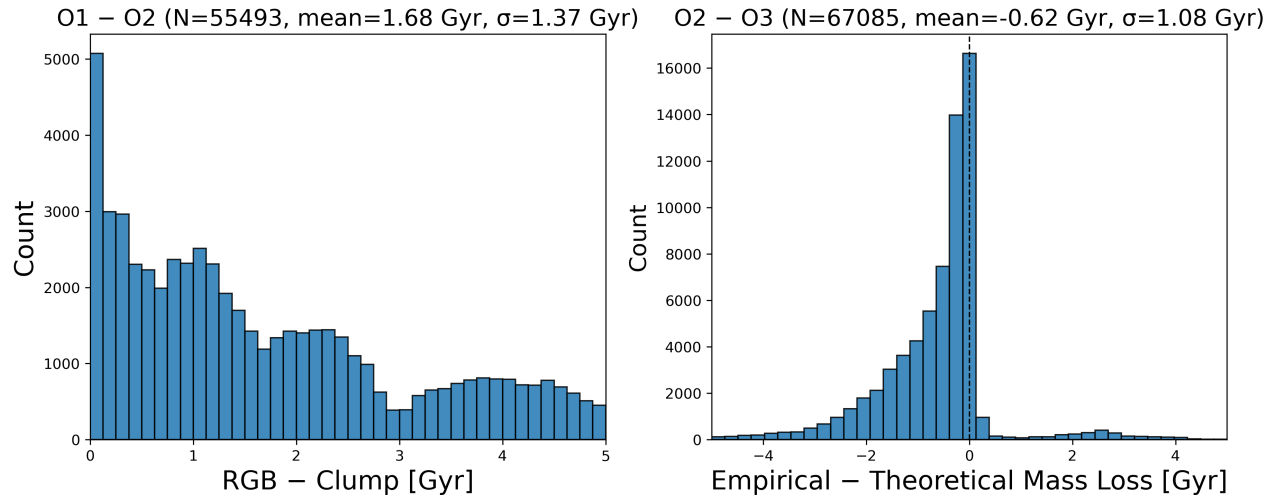

**Figure 6.** Histograms of age differences comparing RGB and clump stars, used to assess how mass-loss prescriptions affect inferred ages. The left panel shows $\Delta{\rm Age} = {\rm Age}_{\rm RGB} - {\rm Age}_{\rm Clump}$, highlighting that changing the assumed evolutionary state can substantially alter the inferred age because of mass loss. The right panel shows $\Delta{\rm Age} = {\rm Age}_{\rm Empirical} - {\rm Age}_{\rm Theoretical}$, where the two different mass-loss schemes yield minimal differences for the majority of clump stars.

We limited the value for mass loss to a maximum of 0.2 and applied a value of 0.05 for the value $\Delta M$ in any star with a mass greater than 1.3 $M_\odot$, coinciding with accepted expectations for mass loss (M. H. Pinsonneault et al. 2025).

For our theoretical prescription, we identified the track closest to the current mass and metallicity of each star in our dataset, and calculated a $\Delta M$ for each of these stars according to the inferred mass expected on the RGB for a star of that mass and metallicity.

Because both mass-loss and evolutionary state identification are uncertain (see Section 5), Table 2 provides ages inferred using our empirically calibrated mass loss, as well as ages calculated with no mass loss, allowing the reader to make alternative choices if desired. Our comparison of ages inferred using the assumption that stars are on the RGB (Option 1) versus on the clump having undergone empirically calibrated mass loss (Option 2) in Figure 6 suggests that the assumed evolutionary state can significantly alter the inferred age. We also show in the right panel of Figure 6 the offset inferred age between assuming an empirical mass-loss rate (Option 2) versus a theoretical mass-loss scaling (Option 3); the choice of which mass-loss parameterization to use generally has a significantly smaller impact age than the question of including mass loss at all, with both schemes often producing similar results.

### *5.2. Evolutionary State Identification*

Because the amount of mass loss that has happened by the RC is significant, and the amount of mass loss that has happened on the early part of the first ascent RGB is insignificant, it is important that we determine which stars are on the RGB and which are in the clump. Unfortunately, in the masses and metallicities of our sample, there is significant overlap between these two populations in temperature and gravity. Therefore, to differentiate them we use the temperatures expected for a first ascent red giant in our models for each star's mass, metallicity, and surface gravity compared to its observed temperature to determine if the temperature is more similar to a red giant of this mass, or more similar to a core-helium-burning clump star. To guide the distinction between RGB and RC stars, we introduce the temperature difference metric $T_{\rm eff,Diff}$, which quantifies how far a star's observed effective temperature deviates from the model prediction for its assumed evolutionary phase. Specifically, for each star, we define $T_{\rm eff,Diff}$, which is $T_{\rm Star} - T_{\rm eff,ref}$ (see J. Bovy et al. 2012; J. Tayar et al. 2017; J. A. Holtzman et al. 2018; M. Vrard et al. 2025, for similar implementations). Our $T_{\rm eff,Diff}$ metric is inherently multidimensional, as $T_{\rm eff,ref}$ is derived from mass, $\log g$, and [M/H]. Thus, $T_{\rm eff,Diff} = T_{\rm eff} - T_{\rm eff,ref}$ naturally incorporates dependencies on $T_{\rm eff}$, $\log g$, [M/H], and mass simultaneously. Figure 7 shows the distribution of $\Delta T_{\rm eff}$ with metallicity (left), the collapsed histogram with our adopted 100 K cutoff separating RGB and RC stars (middle), and the H-R diagram, which highlights the clear separation between the two stellar populations (right). We show in Figure 7 that the distribution of $T_{\rm eff,Diff}$ has two clear peaks at 45 and 208 K. The peak close to zero offset from the expected RGB temperature represents our red giants, while the peak around 200 K represents RC stars, which are expected to be hotter than first ascent giants of the same mass. While there is clearly overlap between the clump and giant populations, we choose to divide them at a temperature offset of 100 K, which is chosen to provide a purer sample of red giant stars. While other studies may adopt different cutoff values depending on their scientific goals, we find that $\Delta T_{\rm eff} = 100$ K provides a suitable balance for our analysis, prioritizing the precision of RGB stellar parameters and ages. We note that varying the temperature of the division by $<15$ K alters the sample distribution by roughly 2%, and we provide ages for each star based on evolutionary state in Table 2.

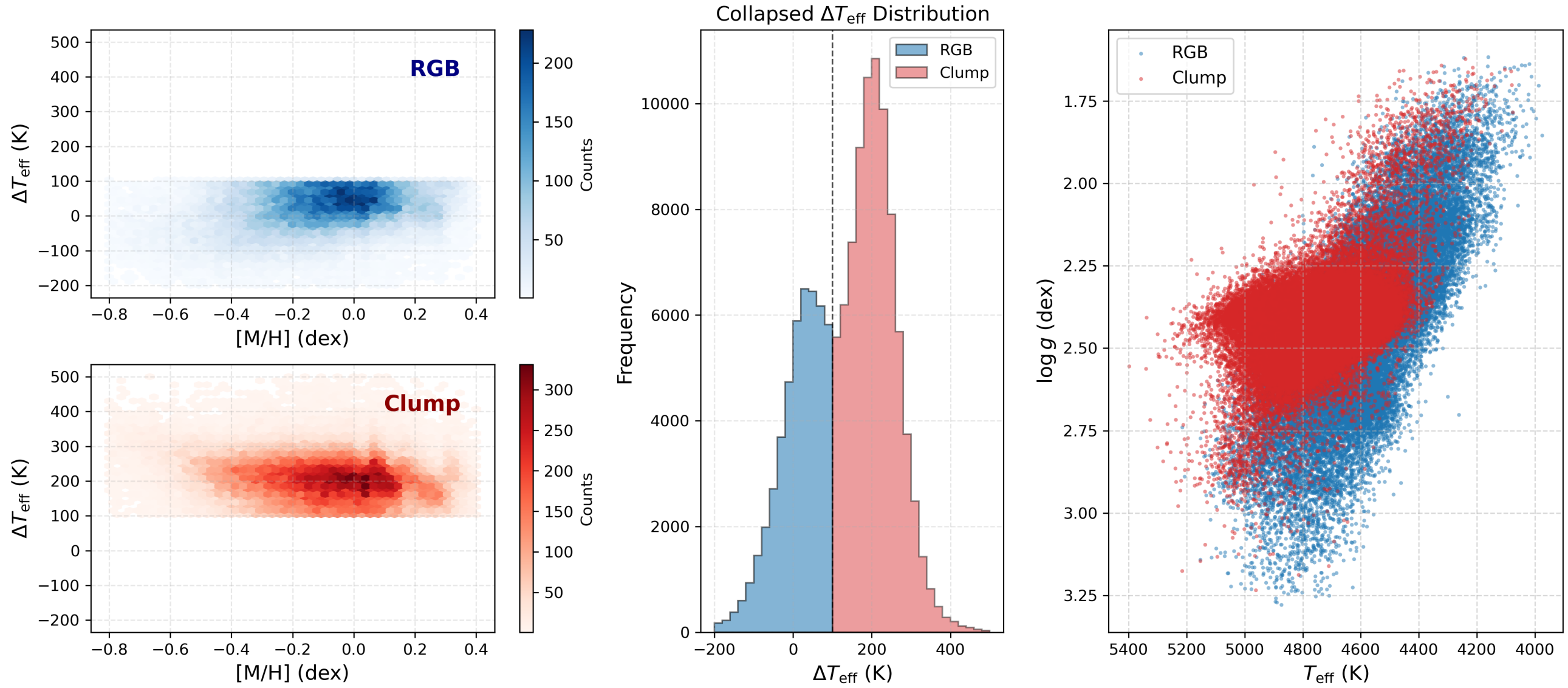

**Figure 7.** Effective temperature differences and classification of evolutionary state. Left: distribution of effective temperature differences ($\Delta T_{\rm eff} = T_{\rm eff,XGB} - T_{\rm eff,ref}$) as a function of metallicity ([M/H]). RGB stars are shown in cyan and RC stars in red, separated by our adopted threshold at $\Delta T_{\rm eff} = 100$ K. Middle: histogram of $\Delta T_{\rm eff}$ for the same sample, with a vertical dashed line at 100 K marking the RGB–RC division. This histogram includes a typical ±50 K uncertainty. Right: H-R diagram showing the two stellar populations: RGB stars (cyan) and RC stars (red). The RC stars cluster into the distinct RC feature at $\log g \approx 2.5$ and $T_{\rm eff} \approx 4750$ K, confirming the population separation.

## 6. Age Distributions

Figure 8 presents a comparison of the asteroseismic age distributions for RGB and clump stars observed by the TESS and Kepler missions (M. H. Pinsonneault et al. 2025). The histograms are divided into four panels, with the top row showing TESS data and the middle row showing Kepler (APOKASC-3) data. Each row is further divided into RGB stars (left) and clump stars (right), with ages limited to 0–13.8 Gyr for plot visibility and to focus on physically plausible stellar ages. To complement these distributions, the lower panel of the figure shows the fractional age uncertainties as a function of seismic mass for RGB (blue) and RC (red) stars in our sample, providing a direct view of the typical error budget and systematic differences between evolutionary states.

Both TESS and Kepler show similar age distributions for RGB and clump stars, with peaks in the age histograms occurring at comparable values. This suggests that the stellar populations observed by both missions share common characteristics. The general agreement between the two datasets indicates that the age determination methods are consistent across missions, despite differences in observational strategies and target selection. This agreement further supports the robustness of the age estimation methods used in our TESS sample.

The TESS sample is significantly larger, with the number of stars in individual bins ranging up to 3000 for RGB stars and 5000 for clump stars. In contrast, the Kepler sample is much smaller, with the $y$-axis limited to 1000 stars for both the RGB and clump populations. This difference in sample size reflects the broader sky coverage of TESS compared to Kepler, which observed a smaller, fixed region of the sky. Kepler's targets are primarily just above the Galactic disk, while TESS includes stars from a wider range of Galactic latitudes, potentially sampling different stellar populations, leading to shifts in the overall distribution. TESS captures more of the sky including the Galactic plane, resulting in a higher percentage of younger stars observed. In contrast, Kepler observes farther away and therefore sees a higher fraction of old stars from the thick disk and halo.

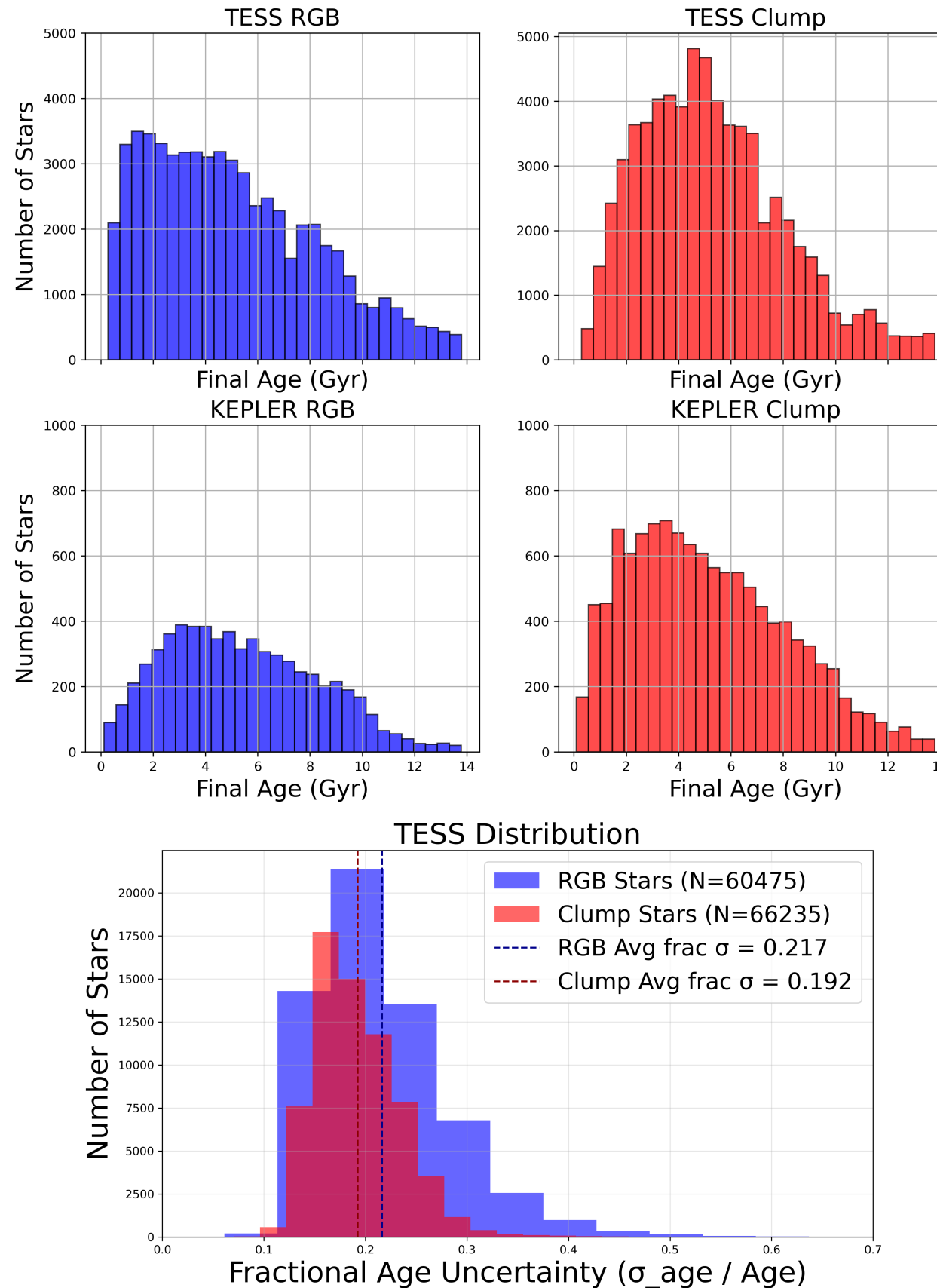

**Figure 8.** Comparison of age distributions for RGB and clump stars observed by TESS (top row) and Kepler (bottom row), divided into four panels with ages limited to the age of the Universe. Both missions show similar age trends, with the TESS samples being significantly larger than Kepler's, while small variances may arise from differences in sky coverage and observation periods. The lower panel shows fractional age uncertainties as a function of seismic mass for RGB (blue) and RC (red) stars, with dashed lines marking the average value for each evolutionary state.

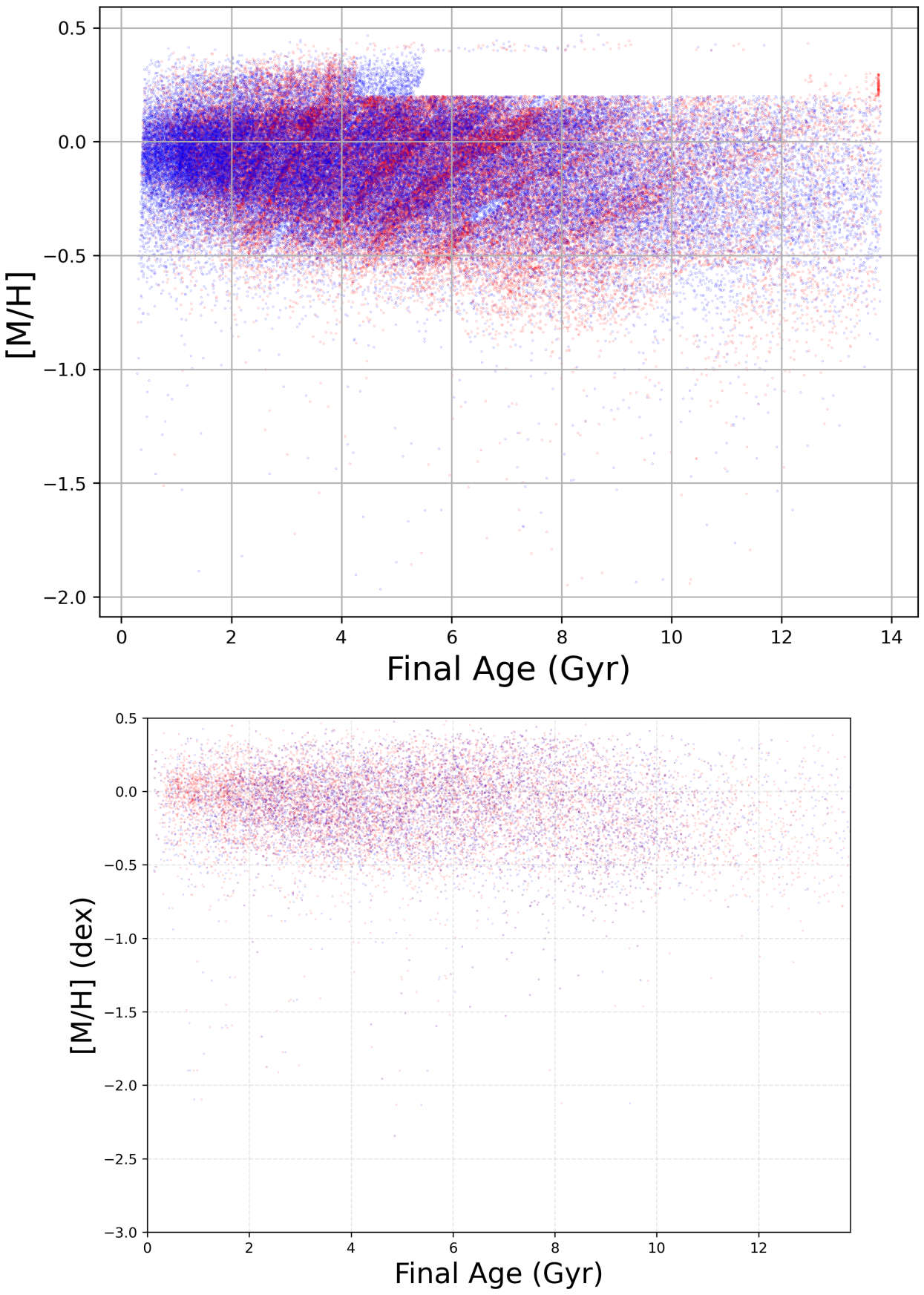

**Figure 9.** Age–metallicity distribution for the entire sample of stars, including the RGB (blue) and RC (red) populations (top). The "striping" pattern arises from interpolation weights biased toward lower-mass tracks and from structure introduced by the mass-loss calculations (see text). The rectangular gap at higher metallicities and old ages reflects missing or incomplete models in our grid. For comparison, the lower panel shows the Kepler/APOKASC age–metallicity distribution for RGB (blue) and RC (red) stars, enabling a visual cross check of trends between datasets.

Both datasets (red giant and clump) show a small fraction (4045/132,794) of stars with ages exceeding the age of the Universe (13.8 Gyr). These outliers may result from uncertainties in age determination or systematic biases in the stellar models used. They also could be the result of binary interactions. Close binaries are more common at low metallicities and higher masses (M. Moe & R. Di Stefano 2017; M. Moe et al. 2019) and some of these old stars, especially in the clump, may represent the remains of stars that have undergone significant envelope stripping (Y. Li et al. 2022). We do not classify any of our stars as potential binary interaction products in our sample, but we note that some of the most extreme outliers may warrant further investigation as candidate binary remnants. Future studies incorporating binary evolution models or additional photometric and spectroscopic diagnostics could provide valuable insights into the nature of these anomalous stars.

Figure 9 shows the age–metallicity distribution for our entire sample of stars, including both RGB and RC populations. The lower panel presents the Kepler/APOKASC age–metallicity distribution for RGB (blue) and RC (red) stars, allowing a direct visual comparison of age trends between the two datasets and illustrating the order-of-magnitude difference in their sample sizes. The plot reveals a distinct "striping" pattern in our sample, which arises from two sources. First, the structure introduced by the mass-loss calculation (see Section 5). Second, the weights assigned during linear interpolation are biased toward the lower-mass track (see Section 4). This clustering highlights the sensitivity of the determined ages to the methodology and grid being used. To mitigate the striping effect, future work could explore increasing the number of interpolation steps or having a more robust mass range with even smaller mass increments.

Additionally, the plot exhibits a rectangular gap at higher metallicities, which arises from the models that failed to run to completion, leading to gaps in the distribution. At cooler effective temperatures, increasing numerical instabilities make stellar models particularly challenging to compute. For this reason, most studies halt evolutionary calculations for stars at roughly 30–50 Gyr, where convergence issues are often found. Some of our evolutionary tracks terminate prematurely, producing an artificial gap in the age distribution around 5 Gyr. Although these convergence issues limit coverage in that specific range, they occur after the primary phases of interest and therefore do not affect the general trends discussed in our analysis. To account for the model-to-model systematic differences in the inferred ages depending on the model grid used (J. Tayar et al. 2022; L. Morales et al. 2025; M. H. Pinsonneault et al. 2025), we have added an additional 10% systematic age uncertainty in quadrature to the observational uncertainties computed using the mass, surface gravity, and metallicity uncertainties. For transparency, all stars in our sample that converged were marked as "0" in the flag column of both tables, and "1" otherwise. Previous work has suggested that this is a representative value for model-to-model absolute age differences for stars in this regime (L. Morales et al. 2025). After conducting crossmatches with wide-binary and open-cluster catalogs (see Figure 8), we find no members of our sample associated with either population. As a result, we cannot perform the age validation in the same manner as Y. Li et al. (2025). Instead, we rely on systematic uncertainties derived from the L. Morales et al. (2025) model and propagated statistical errors. In future work, incorporating larger samples or analysis of specific interesting systems will enable this type of validation to be carried out more robustly. Even with the aforementioned caveats, we feel that our ages are still valuable because of their large number and homogeneously derived nature. Despite these minor challenges, the overall distribution of ages has similar trends to previous studies, and can therefore still be used in future work. We believe it can provide a useful dataset or training sample for a variety of applications.

## 7. Discussion and Conclusions

In this work, we present a sample of ages that can be used for Galactic archeology.

1. We demonstrated that data from TESS and Gaia can be effectively combined to estimate stellar masses for large samples of stars.
2. Using this sample, we constructed a grid of stellar models specifically tailored to its properties. These models incorporate updated physics and are tuned to the range of masses and metallicities observed in our dataset.
3. Leveraging these models, we estimated ages for 132,794 stars with an average uncertainty of <23%. This represents a significant step forward in the size of stellar samples with asteroseismically informed ages available.
4. We quantified both the statistical and systematic uncertainties on our masses and inferred ages. On average, the combined uncertainties were 12% in mass and 23% in age.
5. Our age estimates replicate the trends observed in previous works, validating both our methodology and the model grid used in this study. Our results provide benchmarks for testing galaxy formation models and further our understanding of stellar evolution in low-mass stars.
6. These results highlight the potential of such age estimates for applications in Galactic archeology, potentially facilitating new insights into the formation and evolutionary history of the Milky Way.

This study aims to highlight the value of combining precise observational data with tailored theoretical models. By leveraging TESS's extensive all-sky data and Gaia's precise astrometric and photometric measurements, we addressed two of the major challenges in stellar age determination outlined in our introduction: limited generalizability of stellar samples and systematic discrepancies in age estimates due to model-dependent uncertainties. Specifically, we demonstrated that red giants with seismically inferred mass estimates can produce age estimates with an uncertainty of 23%, providing a useful sample for future Galactic archeology.

The results presented here not only validate the feasibility of extending precise age determinations to a much larger and more representative sample of the Galaxy but also provide critical benchmarks for testing and refining models of the Milky Way's formation and evolution. Our work confirms trends observed in previous studies (e.g., M. Ness et al. 2016; J. T. Mackereth et al. 2018; D. K. Feuillet et al. 2022) and highlights the possibility of utilizing seismically inferred masses and evolutionary states as a tool for age estimation in future work.

Moving forward, these techniques can be applied to explore additional stellar populations, such as those in the Galactic halo or accreted dwarf galaxies, where precise ages could shed light on key questions about their formation and evolution (e.g., A. Helmi et al. 2018; S. K. Grunblatt et al. 2022). Moreover, extending these methods to low-metallicity populations or stars in more extreme environments could help refine our understanding of the interplay between the Galaxy's components, as well as the role of accretion and interactions over cosmic time (W. J. Chaplin et al. 2020).

Future studies may also benefit from integrating asteroseismic techniques with machine learning approaches to further increase the precision and accessibility of stellar ages for even larger datasets (M. Ness et al. 2016; J. T. Mackereth et al. 2019). For example, by building on the machine learning methods validated on Kepler data (A. B. A. Queiroz et al. 2023; A. Stone-Martinez et al. 2024), we can optimize models for the broader and more diverse sample provided by TESS. This integration could also address the challenges of training sample reliance and edge-case inaccuracies, which remain limitations in current age estimation techniques (J. T. Mackereth et al. 2019; J. Tayar et al. 2023).

## Acknowledgments

A.T.T. acknowledges support from the University of Florida CLAS-Scholars and USP-Scholars programs. A.T.T., L.M., and J.T. acknowledge support from 80NSSC22K0812. A..TT. and J.T. acknowledge support from 80NSSC23K0436. We thank Z. Claytor for helpful discussions about `kiauhoku` and M. Joyce for providing her MESA input files.

A.T.T. thanks Georgios and Hatice Theodoridis for helpful discussions and support.

## Appendix
## MESA Inlist Solar Configuration

Inlist_common file used for solar calibrated MESA modeling. Please view documentation for complete process on MESA modeling. This file is available in its entirety on Zenodo at doi:10.5281/zenodo.18050974.

```
!!!!!!!!!!!!!!!!!!!!!!!!!!!!!!!!!!!!!!!!!!!!!!!!!!
!
! authors: A. Theodoridis based on inlist from G. C.
  Cinquegrana & M. Joyce
!
!!!!!!!!!!!!!!!!!!!!!!!!!!!!!!!!!!!!!!!!!!!!!!!!!!
&star_job
  history_columns_file = 'history_columns.list'
  profile_columns_file = 'profile_columns.list'
```

```
    save_model_when_terminate = .true.
    save_model_filename = ’final.model’
    change_initial_net = .true. ! switch nuclear reaction
  network
    new_net_name = ’pp_and_cno_extras.net’
    set_initial_model_number = .true.
    initial_model_number = 0
!  network
    auto_extend_net = .true.
    h_he_net = ’pp_extras.net’ !’basic.net’
    co_net = ’co_burn.net’
    adv_net = ’approx21.net’ !! unclear; do we need
  technetium? do we care?
    num_special_rate_factors = 1
    reaction_for_special_factor(1) = ’r_c12_ag_o16’
    special_rate_factor(1) = 1
    filename_of_special_rate
   (1) = ’r_c12_ag_o16_kunz.txt’
!  opacities
    initial_zfracs = 3 ! for L03 solar scaling
/ !end of star_job namelist
&eos
/ ! end of eos namelist
&kap
    Zbase = 0.0179492d0 ! 0.03d0
    !! AESOPUS needs to be removed when Z = 0.02 or higher
    use_Type2_opacities = .true.
    kap_file_prefix = ’gs98’
    kap_CO_prefix = ’gs98_co’
    ! kap_lowT_prefix = ’AESOPUS’
    ! AESOPUS_filename = ’AESOPUS_GCJul22_allZ.h5’
    kap_lowT_prefix = ’lowT_fa05_gs98’ !! for z= 0.02, 0.05
/ ! end of kap namelist
&controls
initial_mass = 1.0
    initial_y = 0.2725693 !! probably fix
    initial_z = 0.0179492 !0.03 !! loop
    mixing_length_alpha = 1.94 !! from Giulia’s solar
  calibration, so keep it
    MLT_option = ’Henyey’
    num_trace_history_values = 2
    trace_history_value_name(1) = ’rel_E_err’
    trace_history_value_name(2) = ’log_rel_run_E_err’
    !! Artemis change for upper limit Log
    !!log_L_upper_limit = 0
    log_g_lower_limit = -1d99
    log_g_upper_limit = 1d99
    power_h_burn_upper_limit = 1d99
    atm_option = ’T_tau’
    atm_T_tau_relation = ’Eddington’
    atm_T_tau_opacity = ’fixed’
    cool_wind_RGB_scheme = ’Reimers’
    cool_wind_AGB_scheme = ’Blocker’
    RGB_to_AGB_wind_switch = 1d-4
    Reimers_scaling_factor = 0.2d0
    Blocker_scaling_factor = 0.01d0!
  0.01d0 !! eta_Bloecker
    overshoot_scheme(1) = ’exponential’
    overshoot_zone_type(1) = ’any’
    overshoot_zone_loc(1) = ’core’
    overshoot_bdy_loc(1) = ’any’
    overshoot_f(1) = 0.016
    overshoot_f0(1) = 0.004
    overshoot_scheme(2) = ’exponential’
    overshoot_zone_type(2) = ’any’
    overshoot_zone_loc(2) = ’shell’
    overshoot_bdy_loc(2) = ’any’
    overshoot_f(2) = 0.0174
    overshoot_f0(2) = 0.004
min_timestep_limit = 1d-4
 !! testing whether this gets through TACHeB 7/18/23
  energy_eqn_option = ’dedt’ !! copy-
   pasted from 1.3M highZ test suite case 7/20/23
! use_gold2_tolerances = .true.
!! weakening tolerances
relax_use_gold_tolerances = .true.
   use_Ledoux_criterion = .false.
  photo_interval = -1 !100 !50 !! changed to 50
  photo_digits = 5 !3
! if true, write out puls. infos when writing profile
write_profiles_flag = .true. !.true.
write_pulse_data_with_profile = .true. !.true.
pulse_data_format = ’GYRE’
/ ! end of controls namelist
```

## ORCID iDs

Artemis Theano Theodoridis https://orcid.org/0009-0000-7811-0726
Leslie Morales https://orcid.org/0000-0002-1333-8866
Jamie Tayar https://orcid.org/0000-0002-4818-7885

## References

Abdurro’uf, Accetta, K., Aerts, C., et al. 2022, ApJS, 259, 35
Anders, F., Gispert, P., Ratcliffe, B., et al. 2023, A&A, 678, A158
Andrae, R., Rix, H.-W., & Chandra, V. 2023, ApJS, 267, 8
Ash, A. L., Pinsonneault, M. H., Vrard, M., & Zinn, J. C. 2025, ApJ, 979, 135
Bahcall, J. N., Serenelli, A. M., & Basu, S. 2006, ApJS, 165, 400
Bedding, T. R., Mosser, B., Huber, D., et al. 2011, Natur, 471, 608
Berger, T. A., Huber, D., van Saders, J. L., et al. 2020, AJ, 159, 280
Berger, T. A., Schlieder, J. E., & Huber, D. 2023, arXiv:2301.11338
Beyer, A. C., & White, R. J. 2024, ApJ, 973, 28
Bloecker, T. 1995, A&A, 297, 727
Borucki, W. J., Koch, D., Basri, G., et al. 2010, Sci, 327, 977
Bovy, J., Rix, H.-W., & Hogg, D. W. 2012, ApJ, 751, 131
Bovy, J., Rix, H.-W., Green, G. M., Schlafly, E. F., & Finkbeiner, D. P. 2016, ApJ, 818, 130
Brogaard, K., Miglio, A., van Rossem, W. E., Willett, E., & Thomsen, J. S. 2024, A&A, 691, A288
Brown, T. M., Gilliland, R. L., Noyes, R. W., & Ramsey, L. W. 1991, ApJ, 368, 599
Byrom, S., & Tayar, J. 2024, RNAAS, 8, 201
Cantat-Gaudin, T., Anders, F., Castro-Ginard, A., et al. 2020, A&A, 640, A1
Chaplin, W. J., & Miglio, A. 2013, ARA&A, 51, 353
Chaplin, W. J., Serenelli, A. M., Miglio, A., et al. 2020, NatAs, 4, 382
Chen, Y.-Q., Zhao, G., Liu, C., et al. 2015, RAA, 15, 1125
Choi, J., Dotter, A., Conroy, C., et al. 2016, ApJ, 823, 102
Chugunov, A. I., Dewitt, H. E., & Yakovlev, D. G. 2007, PhRvD, 76, 025028
Claytor, Z. R., van Saders, J. L., Santos, Â. R. G., et al. 2020, ApJ, 888, 43
Crawford, C. L., Li, Y., Huber, D., et al. 2025, MNRAS, 542, 3289
Dotter, A. 2016, ApJS, 222, 8
Drimmel, R., Cabrera-Lavers, A., & López-Corredoira, M. 2003, A&A, 409, 205
Eddington, A. S. 1930, MNRAS, 90, 668
El-Badry, K., Rix, H.-W., & Heintz, T. M. 2021, MNRAS, 506, 2269
Ferguson, J. W., Alexander, D. R., Allard, F., et al. 2005, ApJ, 623, 585
Feuillet, D. K., Feltzing, S., Sahlholdt, C., & Bensby, T. 2022, ApJ, 934, 21
Gaia Collaboration 2022, yCat, I/355
Gaia Collaboration, Prusti, T., de Bruijne, J. H. J., et al. 2016, A&A, 595, A1
Gaia Collaboration, Brown, A. G. A., Vallenari, A., et al. 2018, A&A, 616, A1
Gaia Collaboration, Brown, A. G. A., Vallenari, A., et al. 2021, A&A, 649, A1
García Pérez, A. E., Allende Prieto, C., Holtzman, J. A., et al. 2016, AJ, 151, 144
Gilliland, R. L., Brown, T. M., Christensen-Dalsgaard, J., et al. 2010, PASP, 122, 131
Girardi, L. 2016, ARA&A, 54, 95
Godoy-Rivera, D., Pinsonneault, M. H., & Rebull, L. M. 2021, ApJS, 257, 46
Green, G. M., Schlafly, E., Zucker, C., Speagle, J. S., & Finkbeiner, D. 2019, ApJ, 887, 93

Grevesse, N., & Sauval, A. J. 1998, SSRv, 85, 161
Grunblatt, S. K., Saunders, N., Sun, M., et al. 2022, AJ, 163, 120
Hatt, E., Nielsen, M. B., Chaplin, W. J., et al. 2022, yCat, J/A+A/669/A67
Hatt, E. J., Ong, J. M. J., Nielsen, M. B., et al. 2024, MNRAS, 534, 1060
Helmi, A., Babusiaux, C., Koppelman, H. H., et al. 2018, Natur, 563, 85
Henyey, L. G., Forbes, J. E., & Gould, N. L. 1964, ApJ, 139, 306
Holtzman, J. A., Shetrone, M., Johnson, J. A., et al. 2015, AJ, 150, 148
Holtzman, J. A., Hasselquist, S., Shetrone, M., et al. 2018, AJ, 156, 125
Hon, M., Stello, D., & Yu, J. 2018, MNRAS, 476, 3233
Hon, M., Huber, D., Kuszlewicz, J. S., et al. 2021, ApJ, 919, 131
Howell, M., Campbell, S. W., Stello, D., & De Silva, G. M. 2022, MNRAS, 515, 3184
Howell, M., Campbell, S. W., Kalup, C., Stello, D., & De Silva, G. M. 2025, MNRAS, 536, 1389
Huang, C. X., Vanderburg, A., Pál, A., et al. 2020a, RNAAS, 4, 206
Huang, C. X., Vanderburg, A., Ricker, G. R., et al. 2020b, Quick-Look Pipeline (QLP) Data Release, MAST, doi:10.17909/t9-r086-e880
Hubeny, I., & Lanz, T. 2017, arXiv:1706.01859
Hubeny, I., Allende Prieto, C., Osorio, Y., & Lanz, T. 2021, arXiv:2104.02829
Huber, D., Bedding, T. R., Stello, D., et al. 2011, ApJ, 743, 143
Huber, D., Zinn, J., Bojsen-Hansen, M., et al. 2017, ApJ, 844, 102
Iglesias, C. A., & Rogers, F. J. 1996, ApJ, 464, 943
Iorio, G., Belokurov, V., Erkal, D., et al. 2018, MNRAS, 474, 2142
Jermyn, A. S., Bauer, E. B., Schwab, J., et al. 2023, ApJS, 265, 15
Jönsson, H., Allende Prieto, C., Holtzman, J. A., et al. 2018, AJ, 156, 126
Jönsson, H., Holtzman, J. A., Allende Prieto, C., et al. 2020, AJ, 160, 120
Joyce, M., & Chaboyer, B. 2018, ApJ, 864, 99
Joyce, M., Johnson, C. I., Marchetti, T., et al. 2023, ApJ, 946, 28
Joyce, M., Molnár, L., Cinquegrana, G., et al. 2024, ApJ, 971, 186
Kalarickal Ramachandra, S., Bedding, T. R., Huber, D., et al. 2024, in 8th TESS/15th Kepler Asteroseismic Science Consortium Workshop, 124
Kjeldsen, H., & Bedding, T. R. 1995, A&A, 293, 87
Kunz, R., Fey, M., Jaeger, M., et al. 2002, ApJ, 567, 643
Lebreton, Y., Goupil, M. J., & Montalbán, J. 2014, EAS, 65, 99
Leung, H. W., & Bovy, J. 2019, MNRAS, 489, 2079
Li, Y., Bedding, T. R., Murphy, S. J., et al. 2022, NatAs, 6, 673
Li, Y., Bedding, T. R., Huber, D., et al. 2024, ApJ, 974, 77
Li, Y., Huber, D., Ong, J. M. J., et al. 2025, ApJ, 984, 125
Lund, M. N., Silva Aguirre, V., Davies, G. R., et al. 2017, ApJ, 835, 172
Mackereth, J. T., Bovy, J., Schiavon, R. P. & SDSS-IV/APOGEE Collaboration 2018, IAUS, 334, 265
Mackereth, J. T., Bovy, J., Leung, H. W., et al. 2019, MNRAS, 489, 176
Mackereth, J. T., Miglio, A., Elsworth, Y., et al. 2021, MNRAS, 502, 1947
Majewski, S. R., Schiavon, R. P., Frinchaboy, P. M., et al. 2017, AJ, 154, 94
Marasco, C., Tayar, J., & Nidever, D. 2025, ApJ, 986, 144
Marigo, P., & Aringer, B. 2009, A&A, 508, 1539
Marshall, D. J., Robin, A. C., Reylé, C., Schultheis, M., & Picaud, S. 2006, A&A, 453, 635
Miglio, A., Brogaard, K., Stello, D., et al. 2012, MNRAS, 419, 2077
Moe, M., & Di Stefano, R. 2017, ApJS, 230, 15
Moe, M., Kratter, K. M., & Badenes, C. 2019, ApJ, 875, 61
Morales, L., Tayar, J., & Claytor, Z. 2025, ApJ, 986, 229
Nataf, D. M., Schlaufman, K. C., Reggiani, H., & Hahn, I. 2024, ApJ, 976, 87
Nelder, J. A., & Mead, R. 1965, CompJ, 7, 308
Ness, M., Hogg, D. W., Rix, H. W., et al. 2016, ApJ, 823, 114
Paxton, B., Bildsten, L., Dotter, A., et al. 2011, ApJS, 192, 3
Paxton, B., Cantiello, M., Arras, P., et al. 2013, ApJS, 208, 4
Paxton, B., Marchant, P., Schwab, J., et al. 2015, ApJS, 220, 15
Paxton, B., Schwab, J., Bauer, E. B., et al. 2018, ApJS, 234, 34
Paxton, B., Smolec, R., Schwab, J., et al. 2019, ApJS, 243, 10
Pinsonneault, M. H., Elsworth, Y. P., Tayar, J., et al. 2018, ApJS, 239, 32
Pinsonneault, M. H., Zinn, J. C., Tayar, J., et al. 2025, ApJS, 276, 69
Planck Collaboration, Aghanim, N., Akrami, Y., et al. 2020, A&A, 641, A6
Prša, A., Harmanec, P., Torres, G., et al. 2016, AJ, 152, 41
Queiroz, A. B. A., Anders, F., Chiappini, C., et al. 2023, A&A, 673, A155
Reimers, D. 1975, MSRSL, 8, 369
Ricker, G. R., Latham, D. W., Vanderspek, R. K., et al. 2010, AAS Meeting, 215, 450.06
Rogers, F. J., Swenson, F. J., & Iglesias, C. A. 1996, ApJ, 456, 902
Schonhut-Stasik, J., Zinn, J., Stassun, K., et al. 2023, AAS, 242, 120.08
Silva Aguirre, V., Bojsen-Hansen, M., Slumstrup, D., et al. 2018, MNRAS, 475, 5487
Sinha, A., Zasowski, G., Frinchaboy, P., et al. 2024, ApJ, 975, 89
Stassun, K. G., Oelkers, R. J., Paegert, M., et al. 2019, AJ, 158, 138
Stello, D., Bruntt, H., Preston, H., & Buzasi, D. 2008, ApJL, 674, L53
Stello, D., Saunders, N., Grunblatt, S., et al. 2022, MNRAS, 512, 1677
Stokholm, A., Aguirre Børsen-Koch, V., Stello, D., Hon, M., & Reyes, C. 2023, MNRAS, 524, 1634
Stone-Martinez, A., Holtzman, J. A., Imig, J., et al. 2024, AJ, 167, 73
Tailo, M., Milone, A. P., Lagioia, E. P., et al. 2020, MNRAS, 498, 5745
Tailo, M., Corsaro, E., Miglio, A., et al. 2022, A&A, 662, L7
Takeda, G., Ford, E. B., Sills, A., et al. 2007, ApJS, 168, 297
Tayar, J., Claytor, Z. R., Huber, D., & van Saders, J. 2022, ApJ, 927, 31
Tayar, J., Claytor, Z. R., Fox, Q., et al. 2023, RNAAS, 7, 273
Tayar, J., & Joyce, M. 2025, ApJL, 984, L56
Tayar, J., Somers, G., Pinsonneault, M. H., et al. 2017, ApJ, 840, 17
Theodoridis, A. T., & Tayar, J. 2023, RNAAS, 7, 148
Ting, Y.-S. 2025, OJAp, 8, 95
Ulrich, R. K. 1986, ApJL, 306, L37
van Saders, J. L., & Pinsonneault, M. H. 2013, ApJ, 776, 67
Vrard, M., Pinsonneault, M. H., Elsworth, Y., et al. 2025, A&A, 697, A165
Xiang, M., & Rix, H.-W. 2022, Natur, 603, 599
Ying, J. M., Chaboyer, B., Boudreaux, E. M., et al. 2023, AJ, 166, 18
Zhou, J., Bi, S., Yu, J., et al. 2024, ApJS, 271, 17
Zinn, J. C., Pinsonneault, M. H., Huber, D., & Stello, D. 2019, ApJ, 878, 136
Zorec, J., & Royer, F. 2012, A&A, 537, A120